\documentclass[12pt,preprint]{emulateapj}

\usepackage{color}
\usepackage{graphicx}
\usepackage{natbib, aas_macros}
\bibpunct[,]{(}{)}{;}{a}{}{,}
 \usepackage{multirow}
\usepackage{amsmath,amssymb,mathrsfs}
\shortauthors{HAYASHI \& CHIBA.}
\shorttitle{Dark halo structures of dSphs in the MW and M31}

\begin{document}

\title{STRUCTURAL PROPERTIES OF NON-SPHERICAL DARK HALOS IN MILKY~WAY AND ANDROMEDA DWARF SPHEROIDAL GALAXIES}

\author{Kohei~Hayashi\altaffilmark{1} and Masashi~Chiba\altaffilmark{2}}

\altaffiltext{1}{Kavli Institute for the Physics and Mathematics of the Universe (Kavli IPMU, WPI), The University of Tokyo, Chiba 277-8583, Japan \\E-mail: {\it kohei.hayashi@ipmu.jp}}
\altaffiltext{2}{Astronomical Institute, Tohoku University, Aoba-ku, Sendai 980-8578, Japan \\E-mail: {\it chiba@astr.tohoku.ac.jp}}
\begin{abstract}
We investigate the non-spherical density structure of dark halos of the dwarf spheroidal (dSph) galaxies in the Milky Way and Andromeda galaxies based on revised axisymmetric mass models from our previous work. The models we adopt here fully take into account velocity anisotropy of tracer stars confined within a flattened dark halo. Applying our models to the available kinematic data of the 12 bright dSphs, we find that these galaxies associate with, in general, elongated dark halos, even considering the effect of this velocity anisotropy of stars.
We also find that the best-fit parameters, especially for the shapes of dark halos and velocity anisotropy, are susceptible to both the availability of velocity data in the outer regions and the effect of the lack of sample stars in each spatial bin.
Thus, to obtain more realistic limits on dark halo structures, we require photometric and kinematic data over much larger areas in the dSphs than previously explored.
The results obtained from the currently available data suggest that the shapes of dark halos in the dSphs are more elongated than those of $\Lambda$CDM subhalos. This mismatch needs to be solved by theory including baryon components and the associated feedback to dark halos as well as by further observational limits in larger areas of dSphs.
It is also found that more diffuse dark halos may have undergone consecutive star formation history, thereby implying that dark-halo structure plays an important role in star formation activity.
\end{abstract}


\section{INTRODUCTION}

Dwarf spheroidal (dSph) galaxies in the Milky Way (MW) and Andromeda (M31) galaxies are excellent laboratories to shed light on the nature of dark matter because these galaxies are the most dark matter dominated systems,  with total dynamical-light ratios of 10--1000 \citep{Mat1998,Giletal2007,Woletal2010}.
DSph galaxies also have the advantage that individual member stars can be resolved, thus, it is possible to measure very accurate line-of-sight velocities for their member stars.
Therefore, using these high-quality data, we are able to constrain their internal structure of dark halos in light of the currently standard $\Lambda$-dominated Cold Dark Matter ($\Lambda$CDM) models. In particular, the studies of dSphs are important to understand several controversial issues that $\Lambda$CDM models hold on galactic and sub-galactic scales. 

One of the outstanding issues in $\Lambda$CDM models is the so-called ``core-cusp'' problem: $\Lambda$CDM-based $N$-body simulations have predicted strongly cusped profiles in the center of dark halos~\citep{NFW1996,NFW1997,FM1997,Dieetal2008,Ishietal2013}, whereas the dark halos in the observed galaxies, especially dSphs and low surface brightness (LSB) galaxies, suggest a cored or shallower cusped density profiles~\citep{Moo1994,Bur1995,deBetal2001,Giletal2007,deB2010}.
Recent studies of this problem have argued the possibility that a fraction of the observed dSphs and LSBs can actually have cusped density profiles~\citep{Hayetal2004,Stretal2010,Batetal2011,Stretal2014}.
We note that some specific model assumptions (such as spherical symmetry) and degeneracies (such as between velocity anisotropy and mass distribution) yield uncertainties in the model results, thus it is yet unclear if the most or all of the dSphs indeed have cored central density profiles in their dark halos.

\begin{deluxetable*}{ccccccccc}
\label{tb:tab1}
\tablecolumns{9}
\tablewidth{7.0in}
\tablecaption{The Observational Data set for MW and M31 dSph Satellites}
\tablehead{
Object & No. of Stars & $M_{\ast}$ & $M_V$  & $D_{\odot}$ & $r_{\rm half}$ &  $q^{\prime}$  & ${\tau}_{0.7}$\tablenotemark{a}  &   Refeferences\tablenotemark{b} \\
          &                             & ($10^6M_{\odot}$)  &  & (kpc) & (pc) & (axial ratio) & (Gyr) & }
\startdata
MW dSphs & & & & & & & &\\
Carina     & 776   & $0.38$     & $-9.1 \pm 0.5 $       &$106\pm6$     & $241\pm 23$       & $0.67\pm0.05$   & $4.5$    & 1,3,7\\
Fornax    & 2523 & $20.0$     & $-13.4 \pm 0.3 $     & $147\pm12$  & $668\pm 34$       & $0.70\pm0.01$   &  $4.6$   & 1,3,7\\
Sculptor  & 1360 & $2.3  $     & $-11.1 \pm0.5 $       & $86\pm6$     & $260\pm 39$       & $0.68\pm0.03$   &  $12.7$ & 1,3,7 \\
Sextans  & 445   & $0.44$     & $-9.3 \pm0.5 $       & $86\pm4$     & $682\pm 117$      & $0.65\pm0.05$   &  No data & 1,3,7 \\
Draco     & 185   & $0.29$     & $-8.8 \pm0.3 $        & $76\pm6$     & $196\pm 12$        & $0.69\pm0.02$   &  $11.5$  & 2,7 \\
Leo~I      & 328   & $5.5 $     & $-12.0 \pm0.3 $      & $254\pm15$ & $246\pm 19$        & $0.79\pm0.03$   &  $2.7$   & 2,3,7 \\
Leo~II     & 200   & $0.74 $     & $-9.8 \pm0.3 $    & $233\pm14$  & $151\pm 17$        & $0.87\pm0.05$   &  $6.8$  & 1,7 \\
M31 dSphs & & & & & & & &\\
And~I       & 51    & $3.9  $    & $-11.7 \pm0.1 $     &$745\pm24$    & $670\pm 30$        & $0.78\pm0.04$   & $7.6$  & 5,6,7\\
And~II      & 488  & $7.6  $    & $-12.4 \pm0.2 $  & $652\pm18$    & $1230\pm20$       & $0.90\pm0.02$   & $6.2$  & 4,6,7\\
And~III     & 62    & $0.83$    & $-10.0 \pm0.3 $ & $748\pm24$    & $400\pm 30$       & $0.48\pm0.02$   & $8.8$  & 5,6,7 \\
And~V     & 85     & $0.39$     & $-9.1 \pm0.2 $    & $773\pm28$    & $350\pm 20$       & $0.82\pm0.05$   & $10.0$ & 5,6,7 \\
And~VII   & 136   & $9.5  $  & $-12.6 \pm0.3 $    & $762\pm35$    & $770\pm 20$       & $0.87\pm0.04$   & $12.8$ & 5,6,7
\enddata
\tablenotetext{a}{This value is the lookback time at achieving $70\%$ of current stellar mass of dSphs, and is estimated by available data taken from~\citet{Weietal2014}.}
\tablenotetext{b}{References: (1)~\citet{IH1995}; (2)~\citet{Maretal2008}; (3)~\citet{Waletal2009c}; (4)~\citet{Hoetal2012};~(5)~\citet{Toletal2012}; (6)~\citet{MI2006}; (7)~\citet{McC2012}} 
\end{deluxetable*}

Another example of the tension between $\Lambda$CDM predictions and observations is the ``too-big-to-fail'' problem that masses of dark halos associated in the dwarf satellites of the MW are significantly lower than those of the most massive subhalos in an MW-sized halo in the $\Lambda$CDM universe~\citep{Boyetal2011,Boyetal2012}.
This problem may be regarded as a rewriting of the ``core-cusp'' problem, where the presence of a core in the center of a halo tends to lower its mean mass density inside the luminous parts  \citep[e.g.,][]{BD2014,OB2015}.
Thus, to solve both problems simultaneously, a possible solution may rely on a transformation mechanism from a cusped to cored central density, e.g.,  through the effects of baryonic process.
In recent years, high-resolution cosmological $N$-body and hydrodynamical simulations have attempted to alleviate the above issues; inner dark halo profiles at dwarf-galaxy scales can be made cored due to energy feedback from gas outflows driven by star-formation activity of galaxies such as radiation energy from massive stars, stellar winds, and supernova explosions~\citep{Govetal2012,Penetal2012,Garetal2013,BZ2014,DiCetal2014a, DiCetal2014b,Madetal2014,OM2014,Onoetal2015}.  
For instance, using the results from cosmological $\Lambda$CDM simulation of isolated dwarf galaxies that can reproduce observed stellar mass-metallicity relations, metallicity distribution functions, and star-formation histories in the Local Group, \citet{Madetal2014} investigated whether both the core-cusp and too-big-to-fail problems can be mitigated based on supernova feedback. They found that in the simulated dwarfs, the total energy injected by SNe~II exceeds the minimum energy required for the core-cusp transformation at all times. Thus, they concluded that simulated dwarf galaxies can form a cored dark halo and have a maximum circular velocity that is lower than the case for dissipationless dark matter only, because supernova feedback can effectively reduce dark matter density in the inner part of its dark halo. 
In the light of suggestions from these simulations, dark halo structures can be related with the evolution of dwarf galaxies such as star-formation history (SFH) and chemical enrichment.

The shapes of dark halos, especially those of subhalos associated with a host dark halo, provide important information on the dynamical evolution of subhalos, as well as the structure of stellar systems formed within their subhalos.
This is because the shapes of subhalos are generally subject to the strength and frequency of tidal effects from the host halo~\citep{Kuhetal2007} and facilitate appropriate investigations about the validity of some of the assumptions in the mass modeling of stellar kinematics in dwarf galaxies~\citep{Veretal2014}.
$N$-body simulations based on $\Lambda$CDM theory have resolved that dark halos have generally triaxial shapes~\citep{JS2000,JS2002,Alletal2006,Hayetal2007}. 
Also, all of recent $N$-body simulations have predicted that dark subhalos, which are at the low-mass end of a mass function for dark halos as well as galaxies, are not strongly triaxial~\citep{Kuhetal2007,Schetal2012}, but rather statistically oblate, axisymmetric shapes~\citep{Veretal2014}.  
Photometric observations of dSphs in the MW and M31 also show that their light distributions on the sky are actually non-spherical~\citep{IH1995,MI2006}.
Therefore, to describe the internal structures of non-spherical dark halos and stellar systems in dSphs, we should use non-spherical mass models for these galaxies rather than spherical mass models. 

\citet[][hereafter HC12]{HC2012}~constructed axisymmetric mass models, where each of luminous and dark matter distribution has a non-unity axial ratio, and  applied these models to line-of-sight velocity dispersion profiles of six bright dSphs in the MW to constrain the non-spherical structure of their dark halos. They concluded that the shapes of these dark halos are very flattened (axial ratio of dark halo: $Q\sim 0.3-0.4$) for most of the sample dSphs.
We note that their axisymmetric mass models assume the distribution function of stars in the form $f(E,L_z)$, where $E$ and $L_z$ denote the binding energy and angular-momentum component toward the symmetry axis, respectively. Under this simplified assumption, velocity dispersions of stars, $\overline{v^2_R}$ and $\overline{v^2_z}$, in the $R$ and $z$ directions in cylindrical coordinates, respectively, are identical, or in other words, a velocity anisotropy parameter $\beta_z$ defined as $\beta_z=1-\overline{v^2_z}/\overline{v^2_R}$ is zero.
The models are thus limited to address the degeneracy between velocity anisotropy and dark-halo shape as shown by~\citet{Cap2008}, so relaxing this assumption and considering a non-zero $\beta_z$ are both useful to enhance the confidence of the conclusions from axisymmetric mass models.
Thus, in order to obtain a more realistic and reliable mass distribution of the dark halos in dSphs, we employ the axisymmetric mass models that are more generalized than those in HC12 and apply the models to seven MW and five M31 dSphs.

This paper is organized as follows. In \S 2, we briefly describe the photometric and spectroscopic data used for our work and the method of data analysis.
In \S 3, we explain our axisymmetric models for density profiles of stellar and dark halo components based on a Jeans analysis and the method of a maximum likelihood analysis to be applied to the data.
In \S 4 we present the results of our maximum likelihood analysis. Finally, in \S 5 we discuss our results and the implications for dynamical evolution of subhalos and association with star formation activity in dSphs, and present our conclusions.

\section{DATA}
In this section we present the data that is used for fitting our dynamical models.
Table~1 lists the observed properties of seven MW and five M31 dSph galaxies: the number of member stars, the  stellar mass of a galaxy assuming a stellar mass-to-light ratio $(M/L)$ of unity, V-band absolute magnitude, distance from the Sun, projected half-light radius, projected axial ratio of a stellar system, lookback time at achieving $70\%$ of current stellar mass in its SFH (as detailed in Section~5.2), and their references. 
For the kinematic data of their member stars, we use published data as follows. For Carina, Fornax, Sculptor, and Sextans dSphs, we adopt~\citet{Waletal2009a,Waletal2009b}, and for Draco, Leo~I, and Leo~II we adopt~\citet{Kleetal2002}, \citet{Matetal2008}, and \citet{Kocetal2007b}, respectively. 
For the M31 dSphs, we select five dSphs (And~I, And~II, And~III, And~V, and And~VII) such that more than 50 stars which can be used as kinematic data, because for a dSph with less than 50 sample stars it is difficult to evaluate its two-dimensional velocity dispersion maps correctly as described below.
For And~II we use the kinematical data in~\citet{Hoetal2012}, and for the other dSphs in M31, those in~\citet{Toletal2012} are adopted.

The kinematic data that we use here are line-of-sight velocities taken from the above cited papers.
The methodology to remove foreground contamination (i.e. the Galactic dwarf stars) and reliably identify member stars differs in each paper. 
For Carina, Fornax, Sculptor, and Sextans dSphs, their member stars are estimated by `expectation-maximization' method in~\citet{Waletal2009b}. 
For Draco, the separation of the member stars from the Galactic contaminant stars is clearly made, so there is little likelihood of non-Draco stars being included in samples.
Finally, for Leo~I, Leo~II, and selected M31 dSphs,  the method for evaluating membership is basically a simple kinematical selection, i.e., the member stars of those dSphs are determined from their velocity distribution (Leo~I:~\citealt{Matetal2008}, Leo~II:~\citealt{Kocetal2007a}, And~II:~\citealt{Hoetal2012}, the other M31 dSphs:~\citealt{Toletal2012}), which is well distinguished from that of the Galactic foreground stars.

In order to estimate the line-of-sight velocity dispersion profiles, we adopt here the standard approach of using binning profiles. 
To begin, since we suppose an axisymmetric system in this work, we analyze the velocity data by folding up the stellar distribution into the first quadrant in each dSph.
In HC12, they estimated line-of-sight velocity dispersion profiles along three axes: major, minor, and intermediate axes (the last of which is defined at 45$^{\circ}$ from major axis; see Section~3.2 in HC12 for further details).
However, their method does not utilize all of the available member stars in each dSph asa they focus on the data only along these three axes, thus their profiles may not fully reflect all of the data that is actually available for each galaxy.
To avoid this, we utilize all of the velocity data and derive two-dimensional distribution of line-of-sight velocity dispersions as follows.
First, we represent stellar distribution in the sky (of the first quadrant of each dSph) in two-dimensional polar coordinates $(r,\theta)$, where $\theta=0$ is set along the major axis, and then divide this into three areas in increments of $30^{\circ}$ in the direction from $\theta=0^{\circ}$ to $90^{\circ}$.
For convenience, we call these three areas the major axis area, which is defined in the region between $\theta=0^{\circ}$ and $30^{\circ}$, the intermediate axis area with $\theta=30^{\circ}-60^{\circ}$, and the minor axis area with $\theta=60^{\circ}-90^{\circ}$.
Second, for each area, we radially separate stars into bins so that a nearly equal number of stars is contained in each bin: $\sim 100$ stars/bin for Fornax, $\sim 80$ stars/bin for Carina and Sculptor, $\sim 50$ stars/bin for Sextans and And~II, $\sim 20$ stars/bin for Draco, Leo~I, Leo~II, and And~VII, and $10 \sim 15$ stars/bin for And~I, And~III, and And~V.
We thus derive the velocity dispersion maps by using the velocity data of stars contained in each bin.
The above literatures have considered the effect of binary systems on line-of-sight velocity dispersions and concluded that such systems have inflated the velocity dispersion in each bin by at most $\sim 10$~\%, which is small compared to the typical error in the individual velocity dispersion, so that the influence of binary systems in dSphs is in fact negligible.

Filled circles in Figure~\ref{fig:fig1} display the velocity dispersion profiles in three axis areas for 12 dSph satellites. It is found that velocity dispersion profiles show systematic changes from the galactic center and some notable differences between the three areas, so that these information can distinguish the difference in the mass models.
We note that in this figure, each data position along the abscissa corresponds to an average galactocentric distance of stars within each bin in each of the three axis areas for the purpose of demonstration.
But in our analysis, we actually evaluate an average $(r,\theta)$ position of stars within each bin in three axis areas and utilize this averaged value in $(r,\theta)$ when we fit our mass models to the observational data.

\section{MODELS AND ANALYSIS}
\subsection{Jeans equations for axisymmetric systems}
The luminous part of dSphs is not really spherically symmetric, nor are the shapes of dark matter halos predicted by high-resolution $\Lambda$CDM simulations. 
We assume axisymmetry in both stellar and dark-halo components, for which axisymmetric Jeans equations are applied using the velocity dispersion components of stars, $(\overline{v^2_R}, \overline{v^2_{\phi}}, \overline{v^2_z})$, in cylindrical coordinates.
In contrast to HC12, we adopt a non-zero velocity anisotropy parameter, $\beta_z=1-\overline{v^2_z}/\overline{v^2_R}$ and assume $\beta_z={\rm constant}$ as adopted in~\citet{Cap2008} for the sake of simplicity. 
This assumption is actually supported by recent $N$-body simulations; \citet{Veretal2014}~showed that $\Lambda$CDM subhalos have an almost constant $\beta_z$ along the minor axis, and only a weak trend as a function of distance along the major axis.
In this case, the relation between the dark matter halo potential, $\Phi$, and moments of the stellar distribution function is expressed via the Jeans equations:
\begin{equation}
\overline{v^2_z} =  \frac{1}{\nu(R,z)}\int^{\infty}_z \nu\frac{\partial \Phi}{\partial z}dz,
\label{eq:eq1}
\end{equation}
\begin{equation}
\overline{v^2_{\phi}} = \frac{1}{1-\beta_z} \Biggl[ \overline{v^2_z} + \frac{R}{\nu}\frac{\partial(\nu\overline{v^2_z})}{\partial R} \Biggr] + 
R \frac{\partial \Phi}{\partial R},
\label{eq:eq2}
\end{equation}
where $\nu$ describes the three-dimensional stellar density. 
For simplicity, we assume that the density distribution of the stellar system has the same orientation and symmetry as those of the dark matter halo.
These velocity dispersions are provided by the second moments that separate into the contribution of ordered and random motions, as defined by $\overline{v^2} = \sigma^2 + \overline{v}^2$.
Since, these equations indicate intrinsic quantities while we only have projected moments for the dSphs, we derive the projected velocity dispersion from $\overline{v^2_z}$ and $\overline{v^2_{\phi}}$, taking into account the inclination of the object with respect to the observer, following the method given in~\citet{TT2006}.
In the appendix, we describe the detailed model properties of line-of-sight velocity dispersion profiles, especially regarding the effects of velocity anisotropy on the axial ratio of a dark halo, thereby we focus on only the main properties of our model results in what follows.

\subsection{Stellar and Dark Halo Density Models}
Stellar surface densities of dSphs are empirically fit by a Plummer profile~\citep{Plu1911}.
Thus, here we assume this profile and generalize it in the following axisymmetric form,  $\nu(x,y)=L(\pi b^2_{\ast})^{-1} (1+m^{\prime 2}_{\ast}/b^2_{\ast})^{-2}$, where $m^{\prime 2}_{\ast}=x^2+y^2/q^{\prime 2}$, $q^{\prime}$ is the projected axial ratio, and $(x,y)$ are the coordinates aligned with the major and minor axes, respectively.  We estimate $b_{\ast}$ as the projected half-light radius.
One of the benefits in the Plummer profile is that one can recover the analytic three-dimensional density distribution as follows:
\begin{equation}
\nu(R,z) = \frac{3L}{4\pi b^3_{\ast}} \Bigl[1+\frac{m^2_{\ast}}{b^2_{\ast}}\Bigr]^{-5/2},
\label{eq:eq3}
\end{equation}
where $m^2_{\ast} = R^2 + z^2/q^2$, so $\nu$ is constant on ellipses with an axial ratio $q$, and $L$ is the total luminosity.
The intrinsic axial ratio $q$ is related to the projected ratio $q^{\prime}$ and the inclination angle $i$ such as  $q^{\prime 2}=\cos^2i + q^2\sin^2i$, where $i=90^{\circ}$ when a galaxy is edge-on and $i=0^{\circ}$ for face-on. The intrinsic axial ratio can be derived from $q=\sqrt{q^{\prime 2} - \cos^2i}/\sin i$, and thus the inclination angle is limited by $q^{\prime 2} - \cos^2i>0$.

For the dark halo in each dSph, we consider the following density profile
\begin{equation}
 \rho(R,z) = \rho_0 \Bigl(\frac{m}{b_{\rm halo}} \Bigr)^{\alpha}\Bigl[1+\Bigl(\frac{m}{b_{\rm halo}} \Bigr)^2 \Bigr]^{-(\alpha+3)/2},
 \label{eq:eq4}
 \end{equation}
\begin{equation}
m^2=R^2+z^2/Q^2,
\label{eq:eq5}
 \end{equation}
where $\rho_0$ is a scale density, $b_{\rm halo}$ is a scale length in the spatial distribution, and $Q$ is an axial ratio of a dark matter halo.
For simplicity and to focus on only an inner profile of the dark matter halo, we suppose that the density distribution at outer parts is fixed to be $\rho\propto r^{-3}$.
The vantage point of this assumed halo model is that it is straightforward to calculate the gravitational force, enclosed mass, and rotation velocity~\citep[see][]{vanetal1994,BT2008}.

In this work, in order to be determined by fitting to the observed line-of-sight velocity dispersion, we adopt six parameters $(Q, b_{\rm halo}, \rho_0, \beta_z, \alpha, i)$ for each dSph.

\begin{deluxetable*}{cccccccc}
\label{tb:tab2}
\tablecolumns{8}
\tablewidth{7.0in}
\tablecaption{Results of MCMC analysis for 12 dSph galaxies.}
\tablehead{
Galaxy       &      $Q$                  & $b_{\rm halo}$(pc)                & $\rho_0$ $($M$_{\odot}$ pc$^{-3}$)         &  $\beta_z$         & $\alpha$            &$i$ (deg)            &    $L_{\rm prolate}/L_{\rm oblate}$}
\startdata
MW dSphs  &&&&&&&\\
Carina        &   $0.33\pm0.02$    &  $709.7^{+40.6}_{-44.3}$      & $0.107\pm0.006$   &  $-0.05\pm0.06$                 &  $-0.09^{+0.09}_{-0.05}$                &  $87.1^{+2.9}_{-9.1}$      &      $0.0056$ \\ 
Fornax       &   $0.38\pm0.03$    &  $991.1^{+27.0}_{-21.4}$       & $0.086\pm0.003$   &  $-0.17^{+0.16}_{-0.07}$       &  $0.00_{-0.04}$                              &  $90.0_{-10.6}$             &     $0.28$\\ 
Sculptor    &   $0.45\pm0.03$    &  $637.7^{+32.6}_{-26.0}$      & $0.168\pm0.008$   &  $-0.03^{+0.06}_{-0.04}$     &  $0.00_{-0.09}$                              &  $87.8^{+2.2}_{-9.1}$     &      $0.000015$\\ 
Sextans     &   $0.53\pm0.06$    &  $1126.7^{+93.5}_{-74.5}$    & $0.028\pm0.008$   &  $ 0.23^{+0.12}_{-0.18}$       &  $0.00_{-0.10}$                              &  $89.8^{+0.2}_{-12.5}$   &      $0.54$\\ 
Draco        &    $0.40\pm0.05$    &  $590.2^{+46.5}_{-43.9}$     & $0.153\pm0.021$   &   $0.31^{+0.08}_{-0.13}$       &  $-0.86^{+0.11}_{-0.11}$                  &  $75.6^{+14.4}_{-8.8}$   &     $0.21$\\ 
Leo~I        &    $0.86\pm0.10$    &  $581.8^{+33.3}_{-26.1}$      & $0.037\pm0.005$   &  $0.09\pm0.14$                   &  $-1.40^{+0.06}_{-0.08}$                 &  $70.5^{+19.5}_{-7.5}$   &      $0.68$\\ 
Leo~II      &    $0.91\pm0.16$    &  $281.8^{+35.3}_{-30.6}$      & $0.195\pm0.031$    &  $-0.62^{+0.56}_{-1.8}$       &  $0.00_{-0.11}$                                &  $88.8^{+0.2}_{-34.3}$   &     $0.91$\\ 
M31 dSphs  &&&&&&&\\
And~I       &    $2.41^{+0.49}_{-0.39}$     &  $811.3^{+118.1}_{-112.7}$           & $0.037\pm0.009$   &  $0.79^{+0.03}_{-0.05}$     &  $-0.39^{+0.39}_{-0.29}$       &  $90.0_{-34.8}$            &   $0.16$\\
And~II     &    $0.32\pm0.06$               &  $1112.2^{+54.8}_{-55.2}$           & $0.005\pm0.004$    &  $-0.13^{+0.31}_{-0.87}$    &  $-1.03\pm0.09$                     &  $85.2^{+4.8}_{-25.4}$    & $35.1$\\
And~III   &    $0.16\pm0.04$                  &  $796.9^{+90.5}_{-96.8}$           & $0.043\pm0.01$      &  $\leq -0.21$                      &  $-1.43^{+0.14}_{-0.23}$         &  $70.8^{+8.6}_{-3.3}$     &  $0.53$\\
And~V       &    $4.75^{+4.54}_{-1.71}$     &  $369.9^{+35.6}_{-37.2}$           & $0.039\pm0.007$   &  $\leq   0.13$                      &  $-1.33^{+0.21}_{-0.12}$         &  $78.2^{+11.8}_{-14.3}$     &  $0.35$\\
And~VII   &    $0.47\pm0.03$               &  $700.3^{+53.1}_{-44.2}$           & $0.041\pm0.008$    &  $0.57^{+0.09}_{-0.13}$    &  $-1.34^{+0.20}_{-0.16}$         &  $81.2^{+8.8}_{-15.7}$   & $10.2$
\enddata
\end{deluxetable*}

\subsection{Markov-Chain Monte Carlo Analysis}
Our aim is to investigate the properties of a dark matter halo by exploring the best-fitting parameters to the kinematical data of the observed dSph.
To do this, we utilize Markov-chain Monte Carlo (MCMC) techniques with the standard Metropolis-Hasting algorithm~\citep{Metetal1953,Has1970}. 
First we define the likelihood function as
\begin{equation}
L(\bold{M}) = \prod_{i} P_i (\bold{O}_i | \bold{M}).
\label{eq:eqL}
\end{equation}
Here $P_i (\bold{O}_i | \bold{M})$ is the conditional probability of finding observables $\bold{O}_i$ given a set of model parameters $\bold{M}$. 
$\bold{O}_i$ represents the observables for the $i$th source, a description of which is provided later, while $\bold{M}$ represents the halo model parameters described in the previous section: $\bold{M} = (Q, b_{\rm halo}, \rho_0, \beta_z, \alpha, i)$.
To construct the conditional probability $P_i (\bold{O}_i | \bold{M})$, we suppose the conditional probability distribution as follows,
\begin{equation}
P_i (\sigma_{{\rm los},i}| \bold{M}) = \frac{1}{\sqrt{2\pi {\rm Var}(\sigma_{{\rm los},i})}}\exp\Biggl[-\frac{1}{2}\frac{(\sigma_{{\rm los},i}-\sigma_{t,i})^2}{{\rm Var}(\sigma_{{\rm los},i})}\Biggr],
\label{eq:eqP}
\end{equation}
where $\sigma_{{\rm los},i}$ is a line-of-sight velocity dispersion corresponding to observables $\bold{O}_i$, and ${\rm Var}(\sigma_{{\rm los},i})$ is defined as the square of the variance of a line-of-sight velocity dispersion.
$N$ is the number of stars with velocity measurements in the dSph with line-of-sight velocity measurements and $\sigma^2_{t,i}$ is the theoretical velocity dispersion derived from Jeans equations. 
For fitting, the velocity dispersions of our models are evaluated at the average two-dimensional position of each bin associated with the observed stars.

In a practical MCMC method, we calculate the likelihood $L(\bf{M})$ based on the equation~(\ref{eq:eqL}) for the current set of model parameters $\bf{M}$. Then, the next set of $\bf{M}^{\prime}$ is calculated by adding small random fluctuations to the previous $\bf{M}$, and the likelihood $L^{\prime}(\bf{M}^{\prime})$ is calculated.
If $L^{\prime}({\bf M}^{\prime})/ L({\bf M})\geq1$, then the next parameter $\bf{M}^{\prime}$ is accepted. 
If not, we draw a random variable $U$, which has a uniform probability between 0 to 1, and we accept $\bf{M}^{\prime}$ in case of $L^{\prime}({\bf M}^{\prime})/ L({\bf M}) > U$. If $L^{\prime}({\bf M}^{\prime})/ L({\bf M}) \leq U$, the next parameter $\bf{M}^{\prime}$ is rejected and the parameter set remains at the previous one $\bf{M}$.
These procedures are iterated for a large number of trials (at least $\sim 10^5$) because early trials may retain the effects of initial conditions, which is called the initial ``burn-in'' phase.

\section{RESULTS}
In this section, we describe the best-fit dark halo model for each of seven MW dSphs (Carina, Fornax, Sculptor, Sextans, Draco, Leo~I, and Leo~II) and five M31 dSphs (And~I, And~II, And~III, And~V, and And~VII) and present confidence maps between each parameter.
Then, we investigate the impact of photometric observational errors and limited sample volume on the best-fit parameters.

\begin{figure*}[htbp]
\figurenum{1}
 \begin{minipage}{0.5\hsize}
  \begin{center}
   \includegraphics[width=94mm]{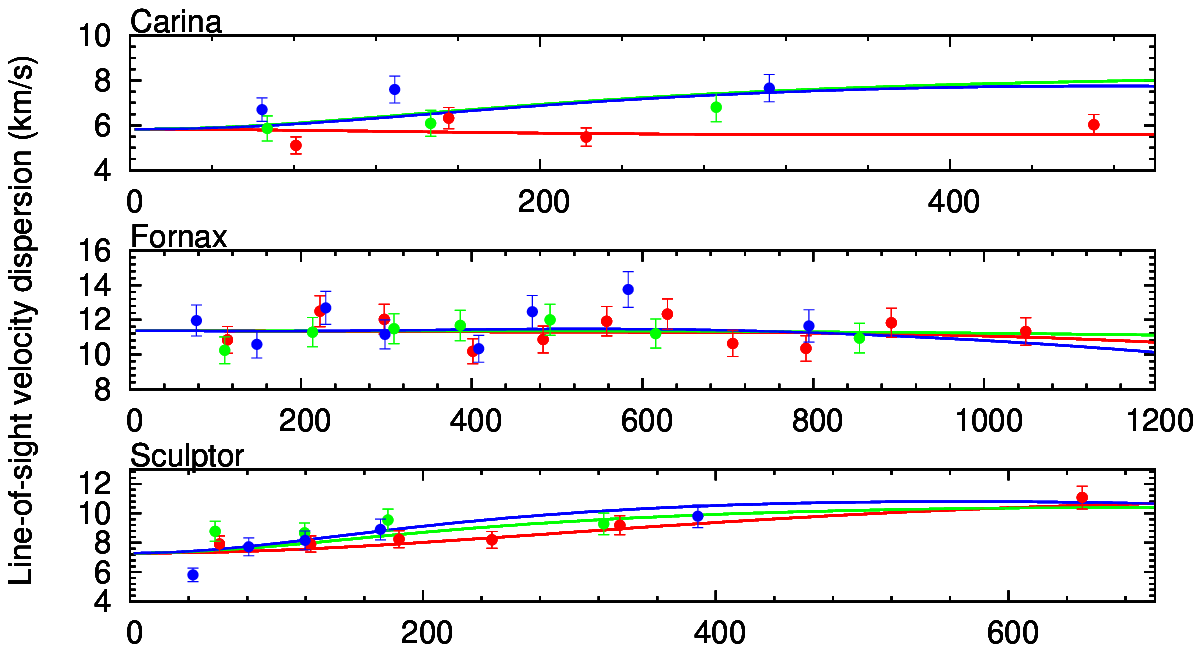}
  \end{center}
 \end{minipage}
 \begin{minipage}{0.5\hsize}
  \begin{center}
   \includegraphics[width=94mm]{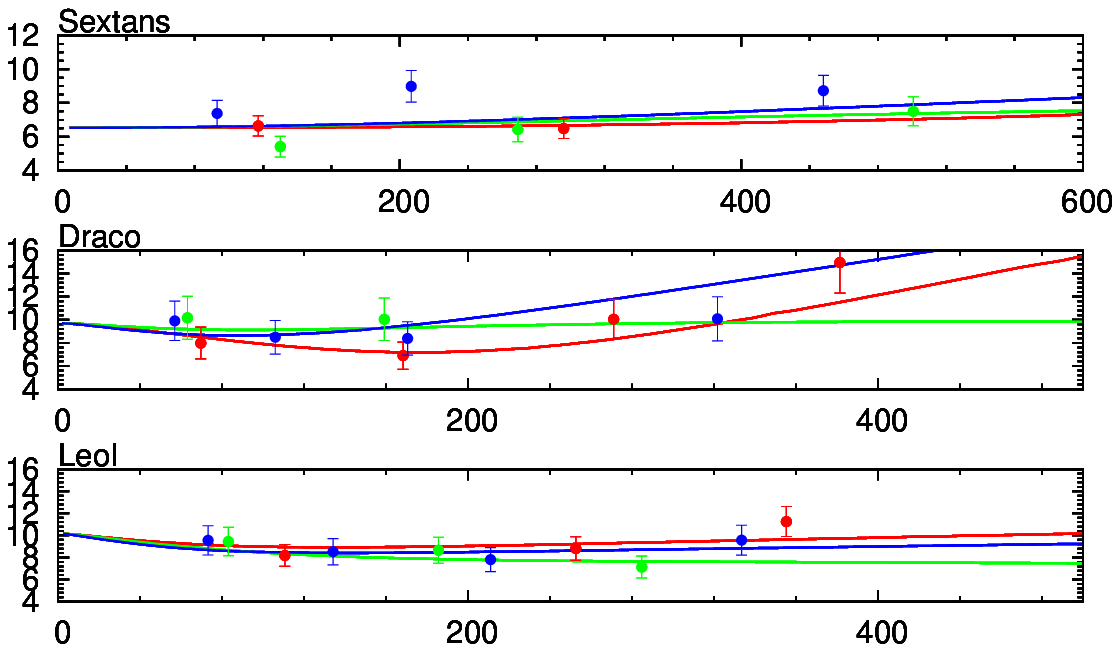}
  \end{center}
 \end{minipage}
  \begin{minipage}{0.5\hsize}
  \begin{center}
   \includegraphics[width=94mm]{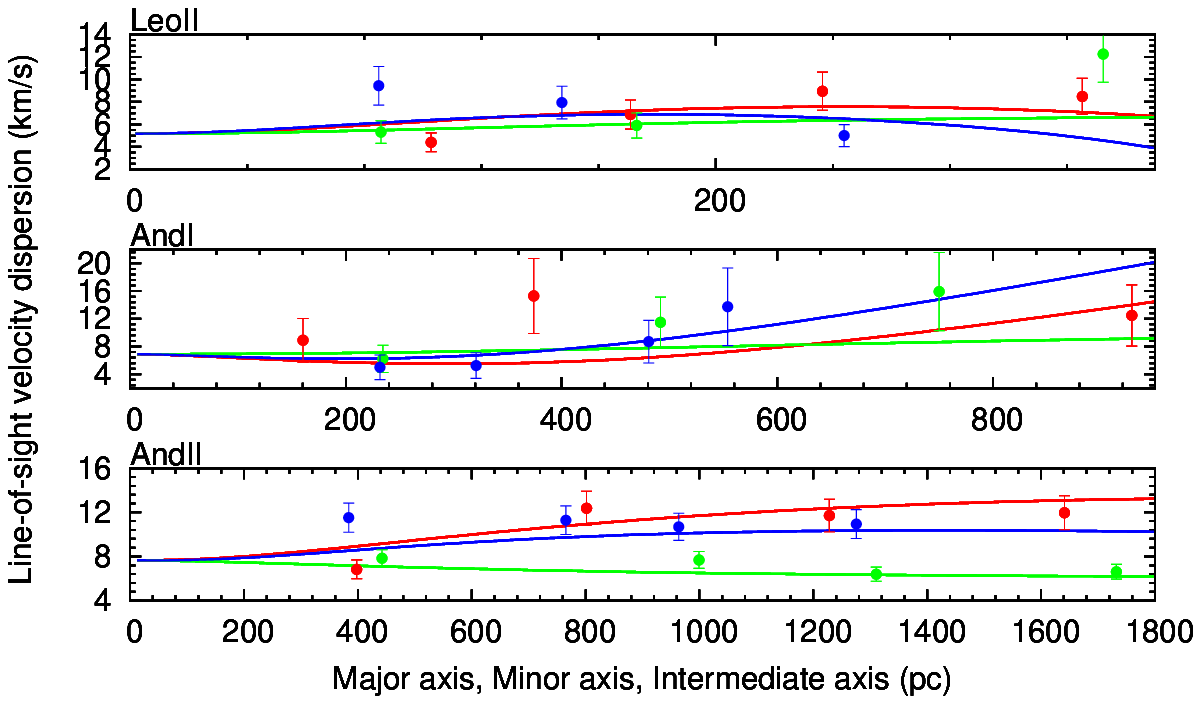}
  \end{center}
 \end{minipage}
 \begin{minipage}{0.5\hsize}
  \begin{center}
   \includegraphics[width=94mm]{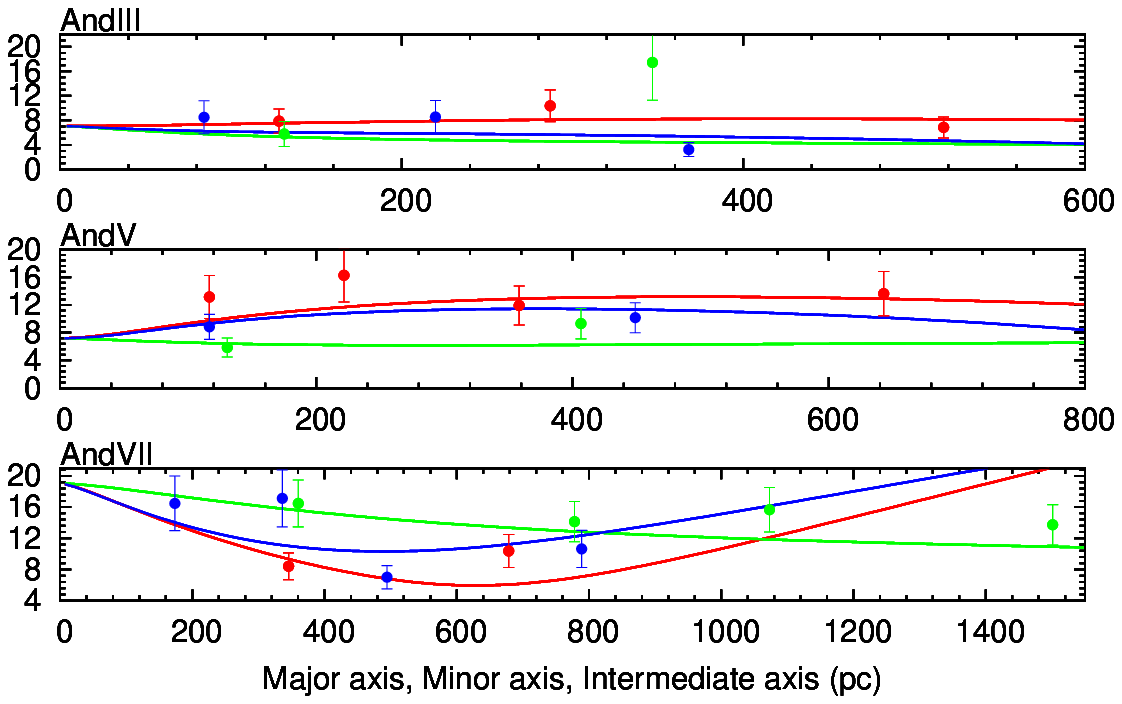}
  \end{center}
 \end{minipage}
\caption{Best-fit profiles of line-of-sight velocity dispersions along major, minor, and intermediate axes. The upper-left three panels are the cases for Carina, Fornax, and Sculptor, while upper-right three are for Sextans, Draco, and Leo~I. The lower-left three panels are for Leo~II, And~I, and And~II, and the lower-right three panels are for And~III, And~V, and And~VII. Red, green, and blue marks denote observed line-of-sight velocity dispersions along major, minor, and intermediate axes, respectively. Red, green, and blue lines are the corresponding best-fit model results along the respective axes.}
  \label{fig:fig1}
 \end{figure*}

\subsection{Best-fit models of dark halos}
We perform the MCMC fitting method for the observed map of line-of-sight velocity dispersions in each dSph mentioned above. 
The best-fit results for each dSph are summarized in Table~2.
Solid lines in Figure~\ref{fig:fig1} show the best-fit line-of-sight velocity dispersion profiles along major, minor, and intermediate axes.
To begin, we investigate if the stellar distribution is either oblate or prolate as judged from the fitting results. 
Column~8 in Table~2 shows the ratio of the most likely value for the prolate case to that for the oblate one.
It is found that for most of the dSphs, the oblate case yields a much better fit than the prolate one, whilst the ratios for some dSphs indicate the prolate one is better, especially for And~II dSph.
Intriguingly, \citet{Hoetal2012} investigated kinematical properties of And~II and found that this dSph is a prolate rotate system, implying that the stellar system itself is prolate.
Thus, our results are consistent with \citet{Hoetal2012}.
In what follows, we use the best-fit parameters in the better fitted stellar distribution. 

Before describing the main results, we should inspect the influence of the use of different bin sizes on our fitting analysis. To do this, we derive the velocity dispersion of three different bin sizes by the same manner as described in Section~2. 
For example, we generate the dispersion maps for Sculptor dSph using three different number densities of stars in each bin: 45, 80, and 110 stars per bin, respectively.
Then we run MCMC analysis for these three types of line-of-sight velocity dispersion data.
Taking the case of Sculptor dSph as an example, we obtain the difference in the best-fit parameters among three different bin sizes is only less than 10~\%. 
In particular, the best-fit shapes of the dark halos appear to take almost the same values: $Q=0.48$, $0.45$, and $0.47$ for these three bin sizes, respectively. 
This is explained by the fact that the velocity dispersion profiles are basically the same among these different types of data sets. Thus, we consider that the difference in the adopted bin size has little effect on the results of our fitting analysis.

\begin{deluxetable*}{cccccccc}
\label{tb:tab3}
\tablecolumns{8}
\tablewidth{5.8in}
\tablecaption{Comparison of Best-fit Parameters Estimated from Full Data Sample with those Estimated from Data Only within Half-light radius}
\tablehead{Object  &  Data  &  $Q$  &  $b_{\rm halo}$(pc)  &  $\rho_0$ $($M$_{\odot}$ pc$^{-3}$)  &  $\beta_z$  & $\alpha$  &  $i$ (deg)}
\startdata
Fornax   &   Data within $r_{\rm half}$   &   $0.83\pm0.04$   &   $708.5^{+23.1}_{-16.7}$   &   $0.101\pm0.004$   &   $0.40\pm0.03$   &  $-0.14\pm0.06$   &  $89.5^{+0.5}_{-10.6}$ \\
              &   Full data                                &   $0.38\pm0.03$   &   $991.1^{+27.0}_{-21.4}$   &   $0.086\pm0.003$   &   $-0.17^{+0.16}_{-0.07}$   &  $0.00_{-0.04}$   &  $90.0_{-10.6}$ \\
&&&&&&&\\
Sculptor &   Data within $r_{\rm half}$   &   $0.73^{+0.07}_{-0.04}$   &   $450.7^{+31.3}_{-17.7}$   &  $0.241\pm0.014$   &   $0.48\pm0.05$   &  $-0.02^{+0.02}_{-0.06}$   &  $89.5^{+0.5}_{-10.6}$ \\
              &   Full data                                &   $0.45\pm0.03$    &  $637.7^{+32.6}_{-26.0}$  &  $0.168\pm0.008$    &   $-0.03^{+0.06}_{-0.04}$  &  $0.00_{-0.09}$   &  $87.8^{+2.2}_{-9.1}$
\enddata 
\end{deluxetable*}

Comparing with the previous results of HC12, which studied the cases for Carina, Fornax, Sculptor, Sextans, Draco, and Leo~I, the values of the best-fitting halo parameters in this work remain roughly the same for these galaxies.
While we expect a finite degeneracy between $Q$ and $\beta_z$ in such a manner that the effect of increasing $\beta_z$ in the fitting of velocity dispersions is similar to that of decreasing $Q$~\citep{Cap2008}, our results as shown in Table~2 indicate that most plausible cases for these dSphs yield $\beta_z\sim0$, so that a flattened dark halo characterized by a small $Q$ is preferred. 
Thus, our conclusion of non-spherical dark halos as the best-fit cases remains unaltered.
Also, it is clear from Column 2 in Table 2 that the shapes of dark halos in the MW and M31 dSphs can be either oblate with $Q<1$ or prolate with $Q>1$. 
This result for a prolate dark halo can be understood as follows, whereas the case for an oblate dark halo has already been explained in Section~4.1 of HC12. 
As shown in Figure~\ref{fig:fig1}, for dSphs with a prolate dark halo like And~I and And~V, the line-of-sight velocity dispersion along the minor axis of these galaxies is relatively small compared with those along the other axes.
In order to reproduce this feature, the best-fit model should yield $Q>1$ so that $\overline{v^2_z}$ decreases in the inner parts.
This is because a larger $Q$ yields a weaker gravitational force in the $z$-direction, thereby giving decreasing $\overline{v^2_z}$ in inner parts, as suggested from the form of equation~\ref{eq:eq1}. 
Also, the $\sigma_{\rm los}$ profile along the minor axis largely reflects $\overline{v^2_z}$ profile, so the observed small $\sigma_{\rm los}$ along the minor axis suggests $Q>1$.
We however note that because of the limitation of the available data volume, $Q$s for these galaxies may be greatly overestimated, as we discuss later.  

\begin{figure*}[t!!!]
\figurenum{2}
 \begin{minipage}{0.45\hsize}
  \begin{center}
   \includegraphics[width=90mm]{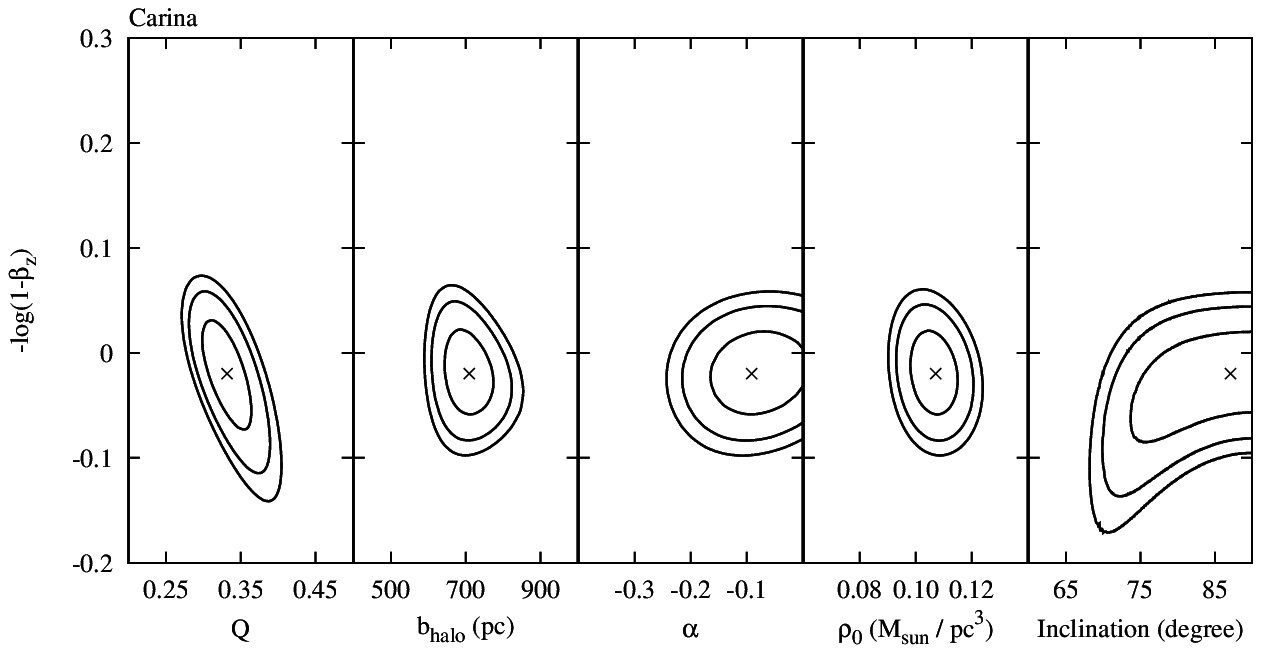}
  \end{center}
 \end{minipage}
 \begin{minipage}{0.5\hsize}
  \begin{center}
   \includegraphics[width=90mm]{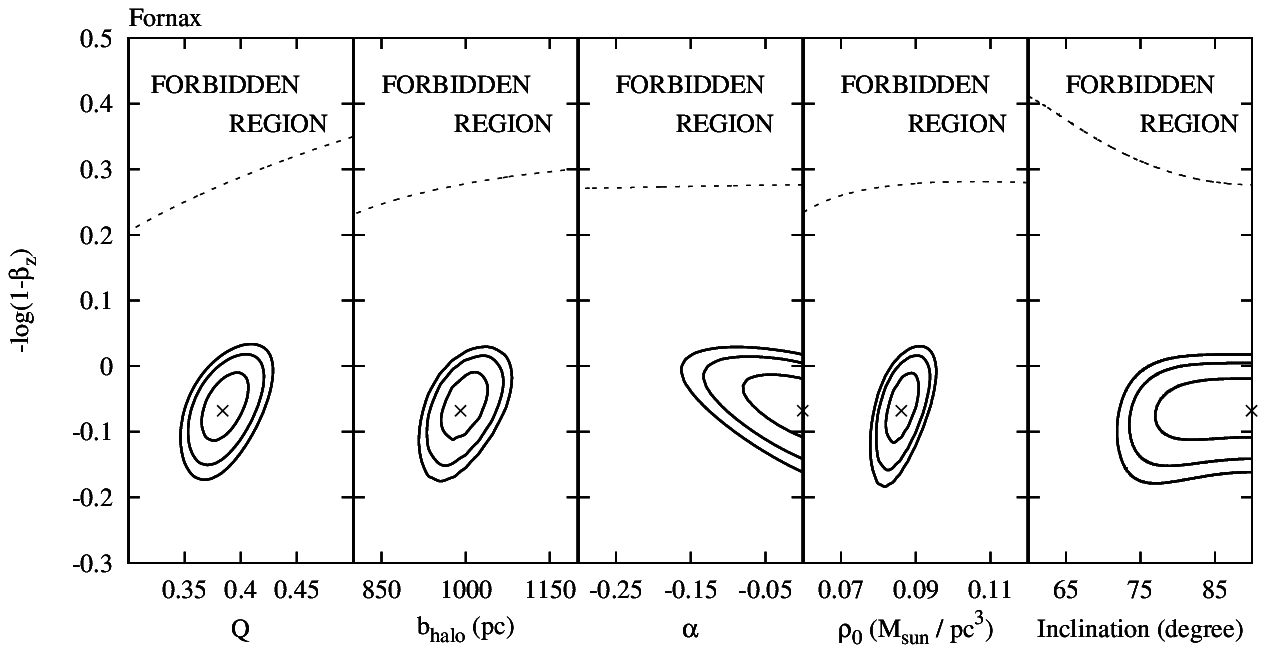}
  \end{center}
 \end{minipage}
  \begin{minipage}{0.45\hsize}
  \begin{center}
   \includegraphics[width=90mm]{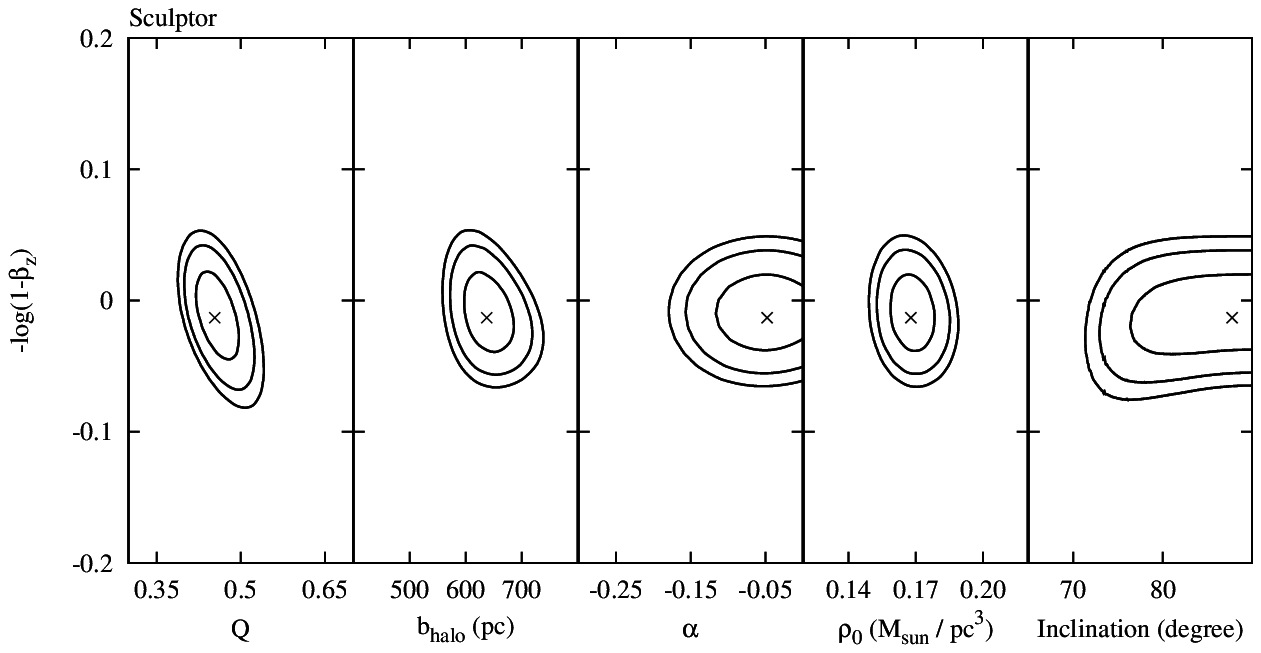}
  \end{center}
 \end{minipage}
 \begin{minipage}{0.5\hsize}
  \begin{center}
   \includegraphics[width=90mm]{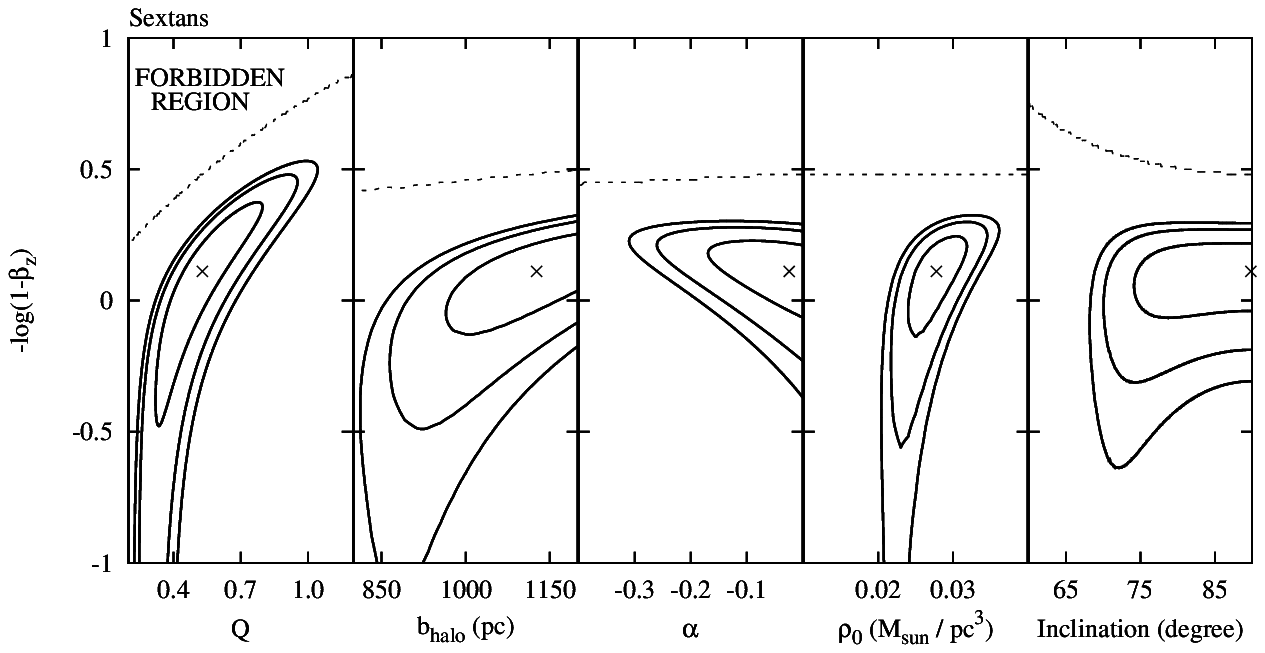}
  \end{center}
 \end{minipage}
  \begin{minipage}{0.45\hsize}
  \begin{center}
   \includegraphics[width=90mm]{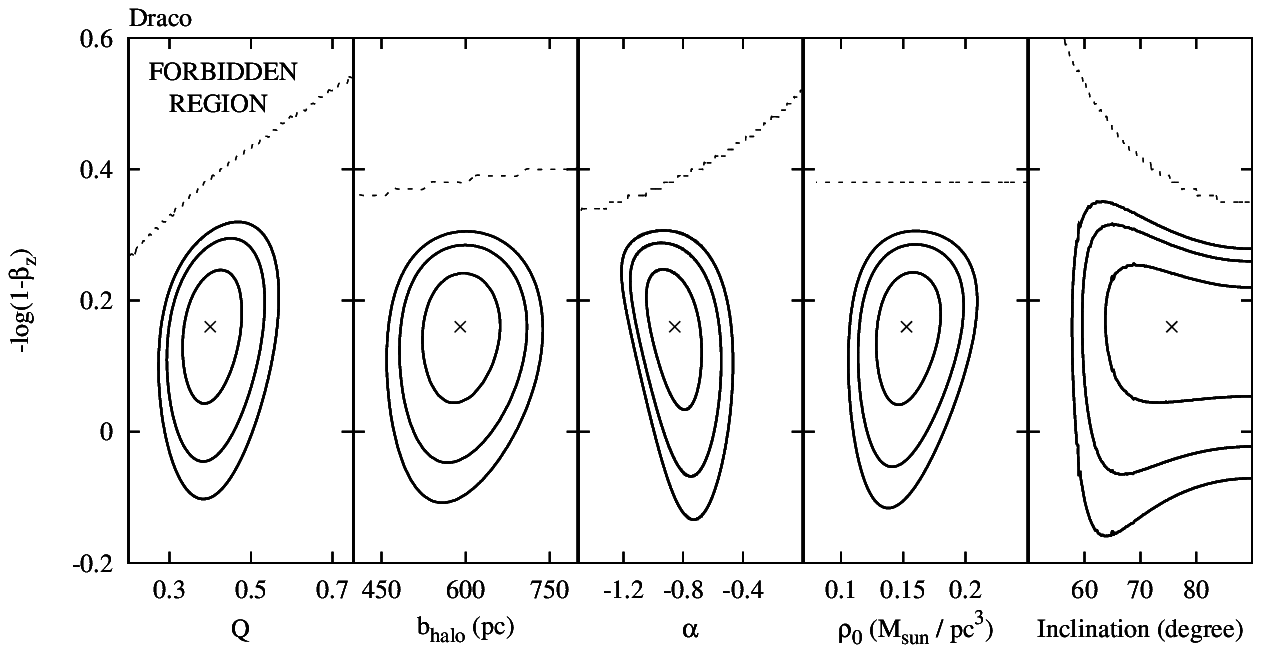}
  \end{center}
 \end{minipage}
 \begin{minipage}{0.58\hsize}
  \begin{center}
   \includegraphics[width=90mm]{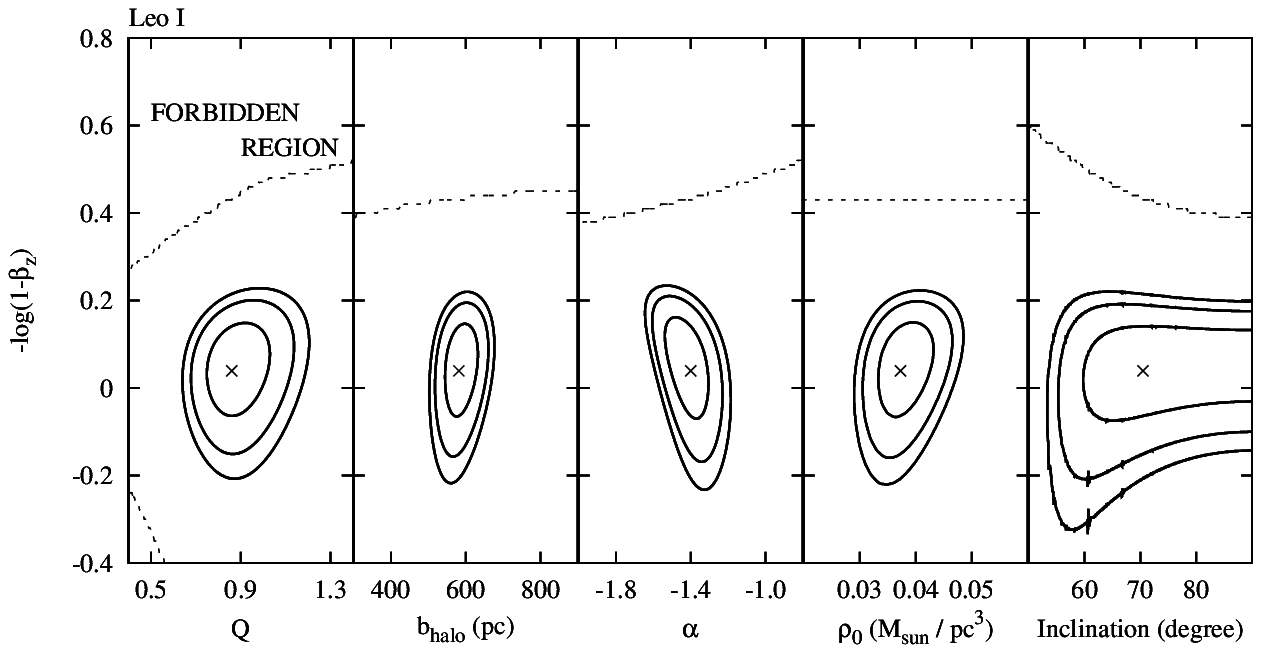}
  \end{center}
 \end{minipage}
 \caption{Likelihood contours for each dark halo parameters against velocity anisotropy, $\beta_z$, for six dSphs. Clockwise from top left, Carina, Fornax, Sextans, Leo~I, Draco, and Sculptor. Contours show 68\% $(1\sigma)$, 95\%, and 99\% confidence levels. The cross point in each panel indicates the best-fit value of each parameter. Dotted line in each panel is a boundary with ``FORBIDDEN REGION'' in which line-of-sight velocity dispersions are unphysical (see the text in Section~4.1 for details).}
  \label{fig:fig2}
 \end{figure*}

\begin{figure*}[t!!]
\figurenum{3}
  \begin{minipage}{0.45\hsize}
  \begin{center}
   \includegraphics[width=90mm]{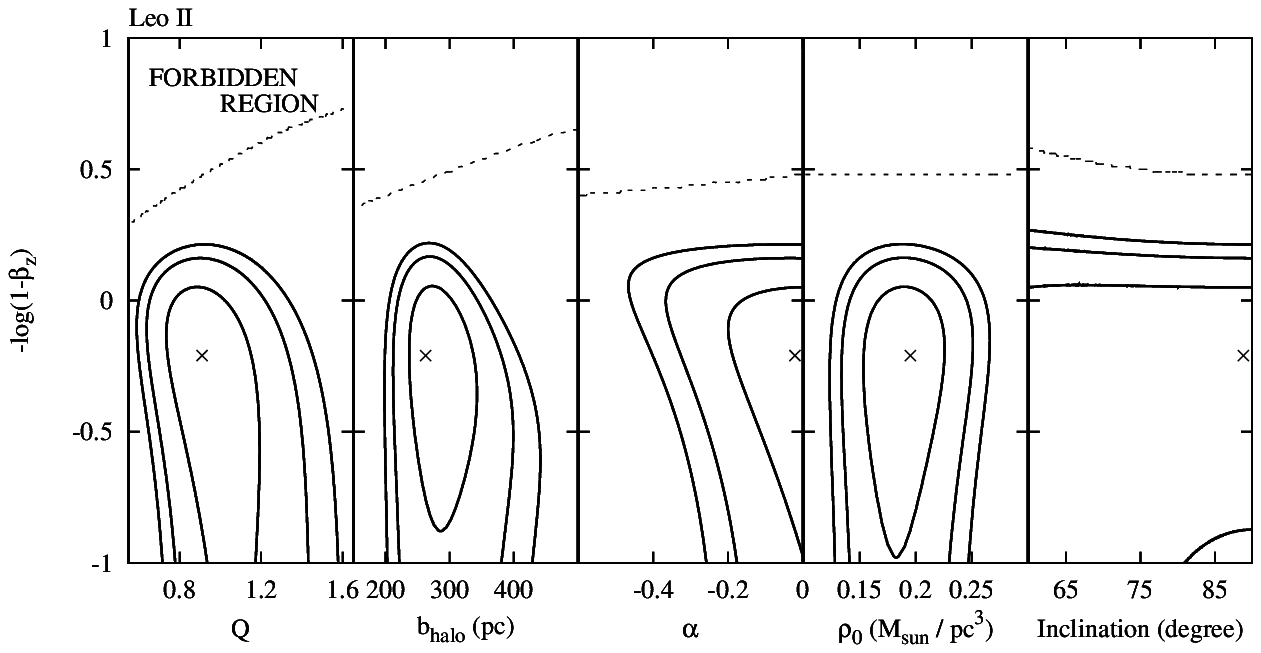}
  \end{center}
 \end{minipage}
 \begin{minipage}{0.5\hsize}
  \begin{center}
   \includegraphics[width=90mm]{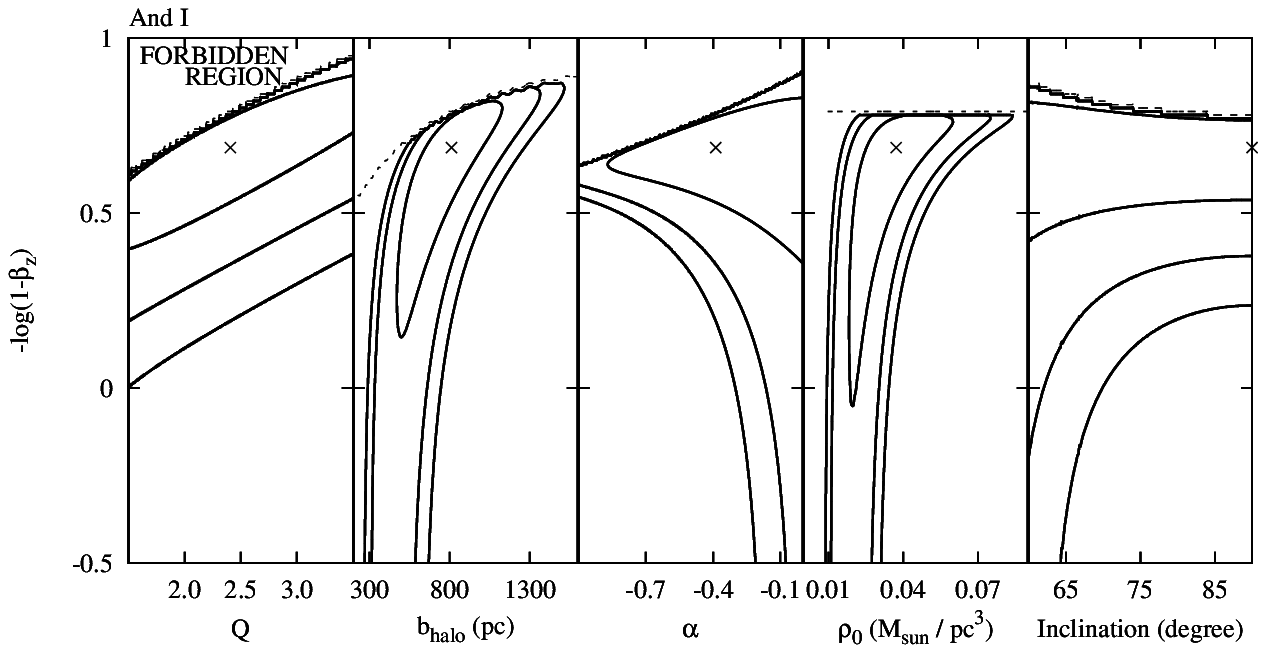}
  \end{center}
 \end{minipage}
   \begin{minipage}{0.45\hsize}
  \begin{center}
   \includegraphics[width=90mm]{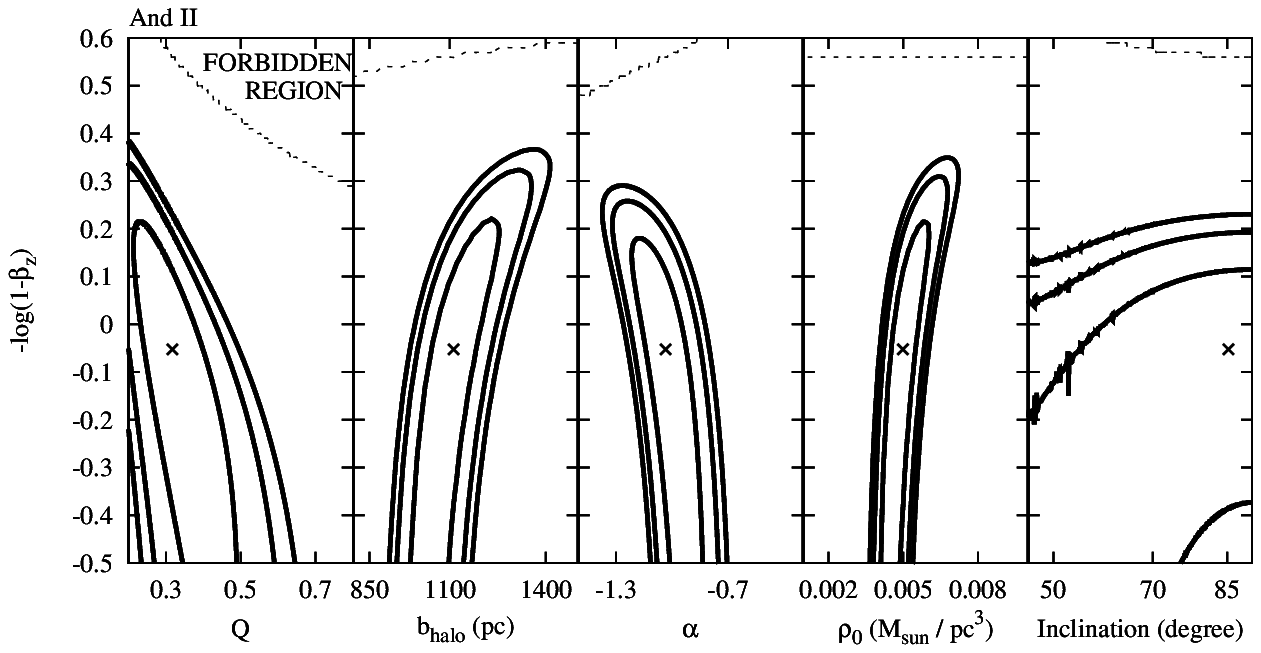}
  \end{center}
 \end{minipage}
 \begin{minipage}{0.5\hsize}
  \begin{center}
   \includegraphics[width=90mm]{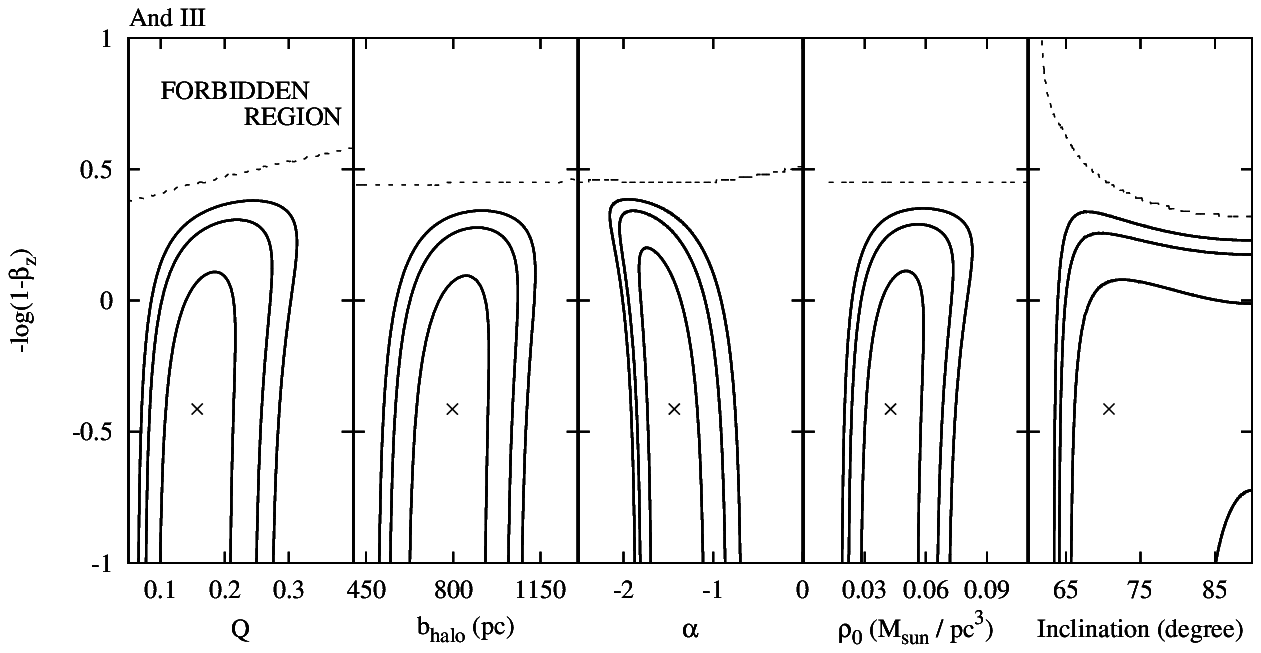}
  \end{center}
 \end{minipage}
  \begin{minipage}{0.45\hsize}
  \begin{center}
   \includegraphics[width=90mm]{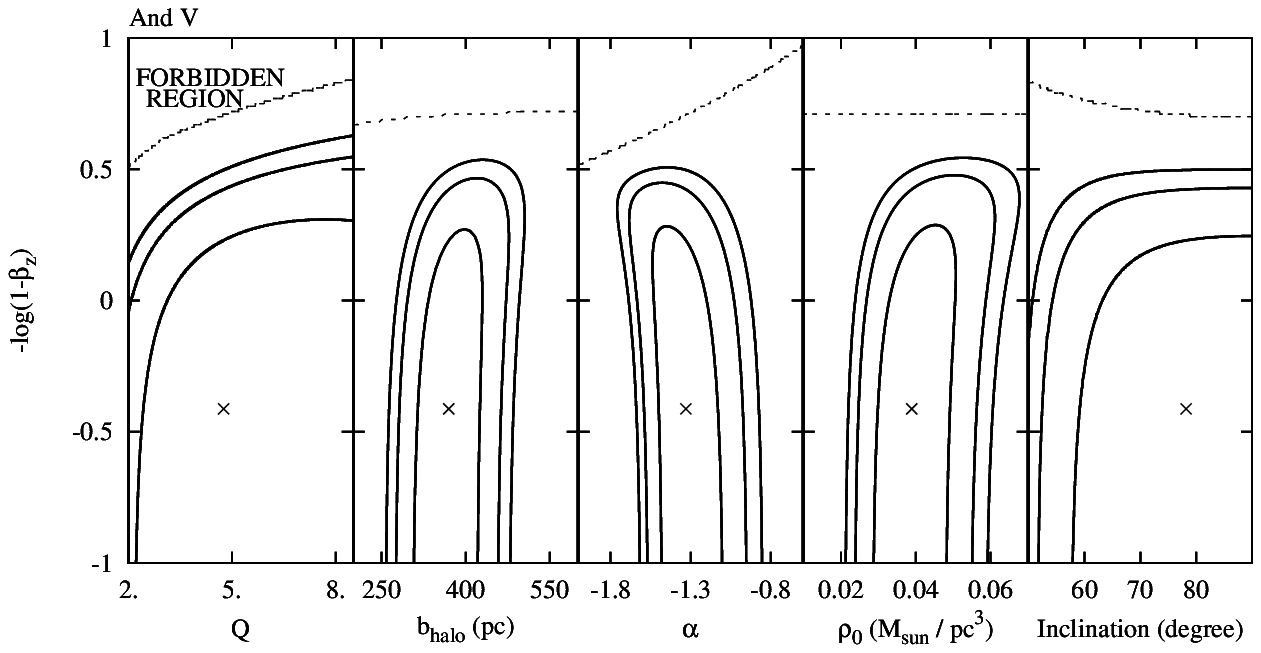}
  \end{center}
 \end{minipage}
  \begin{minipage}{0.58\hsize}
  \begin{center}
   \includegraphics[width=90mm]{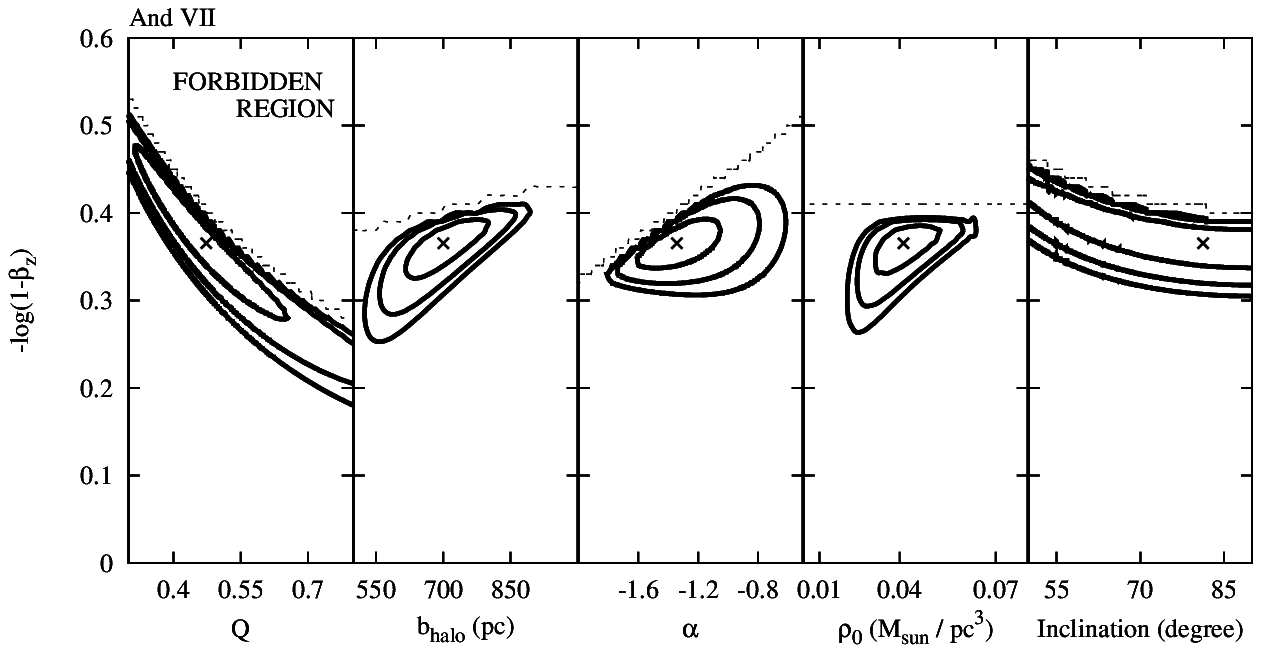}
  \end{center}
 \end{minipage}
 \caption{The same as Figure~\ref{fig:fig2} but for the other dSphs. Clockwise from top left, Leo~II, And~I, And~III, And~VII, And~V, and And~II. }
  \label{fig:fig3}
 \end{figure*}

We investigate other degeneracies in the model fitting for determining these six parameters $(Q, b_{\rm halo}, \rho_0, \beta_z, \alpha, i)$, especially the relation between $\beta_z$ and the other parameters.
In Figure~\ref{fig:fig2} and~\ref{fig:fig3},  we present 68~\% ($1\sigma$), 95~\%, and 99~\% confidence levels of contours in the two-dimensional plane of $Q-\beta_z$, $\rho_0-\beta_z$,  $\alpha-\beta_z$, $b_{\rm halo}-\beta_z$, and $i-\beta_z$ for all sample dSphs: Carina, Fornax, Sculptor, Sextans, Draco, and Leo~I are shown in Figure~\ref{fig:fig2}, and Leo~II, And~I, And~II, And~III, And~V, and And~VII are shown in Figure~\ref{fig:fig3}.
``FORBIDDEN REGION'' indicates the region where line-of-sight velocity dispersions are unphysical, namely the square of $\sigma_{\rm los}$ has negative values in extreme cases such as a very flattened dark halo or very large $\beta_z$.
From these figures, $\beta_z$ appears to have little degeneracies with the other parameters.
In particular, the concerned degeneracy between $Q$ and $\beta_z$ for the  bright dSphs having the large number of member stars turns out to be rather weak.
Whilst this degeneracy certainly exists for line-of-sight velocity dispersion profiles along the major axis, as described in the appendix, it is broken along the minor axis; velocity dispersion profiles along this axis are actually more sensitive to $Q$ than $\beta_z$.
This break of the degeneracy along the minor axis allows us to determine $Q$, provided the sufficient number of stars is available along this axis, as guaranteed for bright dSphs.
In this respect, it is worth noting that for most of Andromeda's satellites, as well as Leo~II, which is very faint, and Sextans, for which global luminosity distribution is yet uncertain because of its very large apparent size, it is impossible to obtain a convergence in $\beta_z$ due to a lack of data sample.
Similarly, for an inclination angle, $i$, the data in the sample dSphs can be reproduced in a wide parameter range of $i$, since the change of $i$ has little influence on line-of-sight velocity dispersion profiles compared to the change of the other parameters. 
Thus, with available data alone it is difficult to determine the best-fit values for both $\beta_z$ and $i$.

\begin{deluxetable*}{cccccccc}
\label{tb:tab4}
\tablecolumns{8}
\tablewidth{6.7in}
\tablecaption{Best-fit Parameters for Considering Photometric Measurement Errors}
\tablehead{Object  &  Measurement Error  &$Q$  &  $b_{\rm halo}$(pc)  &  $\rho_0$ $($M$_{\odot}$ pc$^{-3}$)  &  $\beta_z$  & $\alpha$  &  $i$ (deg)}
\startdata
Fornax   &   {\bf Half-light radius}     &&&&&&\\    
              & Upper limit   &   $0.39\pm0.02$   &   $941.3^{+26.2}_{-17.8}$   &   $0.087\pm0.005$   &   $-0.13\pm0.08$   &  $0.00_{-0.04}$   &  $90.0_{-10.5}$ \\
             &  Lower limit  &   $0.40\pm0.02$   &   $924.7^{+23.5}_{-20.2}$   &   $0.094\pm0.006$   &   $-0.10\pm0.09$   &  $-0.04\pm0.04$  &  $88.6^{+1.4}_{-8.5}$ \\

              &   {\bf Projected  axial ratio}  &&&&&&\\
              & Upper limit   &   $0.40\pm0.02$   &   $887.9^{+28.1}_{-14.9}$   &   $0.095\pm0.004$   &   $-0.16^{+0.10}_{-0.06}$   &  $-0.01^{+0.01}_{-0.03}$   &  $90.0_{-12.1}$ \\
              &  Lower limit   &  $0.37\pm0.02$   &   $1013.9^{+22.8}_{-25.7}$ &   $0.088\pm0.004$   &   $-0.09^{+0.06}_{-0.13}$   &  $0.00_{-0.05}$   &  $90.0_{-10.8}$ \\
&&&&&&&\\
Sextans &  {\bf  Half-light radius}  &&&&&&\\
              & Upper limit   &   $0.69\pm0.08$   &   $1110.6^{+95.1}_{-70.1}$   &   $0.024\pm0.002$   &   $0.37\pm0.11$   &  $0.00_{-0.09}$   &  $90.0_{-12.0}$ \\
              &  Lower limit  &   $0.36\pm0.05$   &   $1410.1^{+161.1}_{-101.1}$ &   $0.026\pm0.003$   &   $\leq 0.16$   &  $-0.10^{+0.08}_{-0.06}$   &  $89.7^{+0.3}_{-12.1}$ \\
              &  {\bf Projected  axial ratio}  &&&&&&\\
              & Upper limit   &   $0.78\pm0.08$   &   $896.7^{+60.7}_{-52.7}$   &   $0.030\pm0.003$   &   $0.38^{+0.07}_{-0.10}$   &  $-0.10^{+0.10}_{-0.08}$   &  $89.8^{+0.2}_{-14.8}$ \\
             & Lower limit    &   $0.48\pm0.06$   &   $1047.7^{+74.6}_{-68.1}$   &   $0.033\pm0.003$   &   $0.29^{+0.11}_{-0.13}$   &  $-0.02^{+0.02}_{-0.08}$   &  $89.5^{+0.5}_{-11.1}$ \\
&&&&&&&\\
And~II   &  {\bf Distance}  &&&&&&\\
               & Upper limit   &   $0.51\pm0.03$   &   $2886.1^{+165.8}_{-110.4}$   &   $0.0085\pm0.002$   &   $0.17\pm0.08$   &  $-0.01^{+0.01}_{-0.07}$   &  $89.8^{+0.2}_{-12.1}$ \\
               &  Lower limit  &   $0.53\pm0.03$   &   $2818.4^{+147.6}_{-150.5}$   &   $0.0093\pm0.002$   &   $0.28^{+0.04}_{-0.07}$   &  $0.00_{-0.042}$   &  $89.7^{+0.3}_{-12.5}$ \\
And~VII &  {\bf Distance}  &&&&&&\\
               & Upper limit   &   $1.87^{+0.45}_{-0.30}$   &   $411.5^{+33.1}_{-26.2}$   &   $0.078\pm0.012$   &   $-0.03^{+0.35}_{-0.54}$   &  $-0.76^{+0.27}_{-0.31}$   &  $70.0^{+20.0}_{-14.1}$ \\
              & Lower limit    &   $1.61^{+0.38}_{-0.27}$    &   $480.5^{+37.5}_{-32.7}$ &   $0.068\pm0.01$     &   $-0.05^{+0.33}_{-0.76}$   &  $-0.61^{+0.24}_{-0.30}$   &  $70.1^{+19.9}_{-14.6}$
\enddata 
\end{deluxetable*}

From our best-fit results, the inclination angles for all dSphs are over 70$^{\circ}$, that is, all stellar systems are nearly edge-on. One of the possible reasons for this result is the limitation of an inclination angle from the relation, $q^{\prime 2} - \cos^2 i > 0$. This inequality suggests that for a given projected axial ratio, $q^{\prime}$, an inclination angle is restricted to be larger than $\cos^{-1}q^{\prime}$. It is also worth noting that in M31's dSphs for which only a small number of spectroscopic data is available, the confidential intervals of $i$ cover rather a large part of parameter space, in contrast to the MW dSphs being confined in high inclination angles.

\subsection{Central density profiles of dark halos}
In the following, we focus on central density profiles of dark matter halos for the sample dSphs.
The question of which central density profiles, cusped or cored, observed dSphs have has been ambiguous, and thus launched the debate known as a core-cusp problem.
\citet{Giletal2007}~gave a review of studies on the inner slope of dSphs' dark halo profile based on spherical Jeans modeling and concluded that cored profiles are preferred for all classical dSphs in the MW.
Recently, applying the relation between the half-light radii and masses within these radii, \citet{WP2011}~inferred the central density profiles in Fornax and Sculptor.
They suggested that Navarro-Frenk-White (NFW)-like cuspy profiles can be ruled out at significance levels $\gtrsim 96$\% and $\gtrsim 99$\% for Fornax and Sculptor, respectively.
Moreover, \citet{Amoetal2013}~and~\citet{AE2012} applied the projected virial theorem to the multiple stellar components for Fornax and Sculptor and argued that this modeling disfavors cuspy profiles in these dSphs.
In contrast, using the axisymmetric Schwarzschild method,~\citet{Jaretal2013} and~\citet{JG2013} applied their axisymmetric stellar models to MW dSphs. They concluded that the dark halo profiles in the dSphs is similar to NFW profiles.
\citet{Stretal2014}~assumed the distribution function of the stellar system embedded in a spherical NFW dark halo, and showed that this distribution function reproduces the line-of-sight velocity dispersions and surface stellar densities of two subcomponents in the Sculptor dSph.
On the other hand, \citet{Waletal2009c}~and~\citet{BH2013} concluded it difficult to constrain an inner profile of a dark halo with current data quantity and quality.  

We point out that all of the previous studies have assumed a spherical dark halo. 
Thus, our non-spherical dark halo models are expected to set useful constraints on central profiles of dark halos. 
From our fitting results, we find that not all of the dark halos in the sample dSphs have a cored central density profile; while most of the dSphs indicate constant density profiles or shallower cusps, the Draco, Leo~I, And~II, And~III, And~V, and And~VII dSphs show a steep inner density slope, $\alpha$, with $-0.86^{+0.11}_{-0.11}$, $-1.40^{+0.06}_{-0.08}$, $-1.03^{+0.09}_{-0.09}$, $-1.43^{+0.14}_{-0.23}$, $-1.33^{+0.21}_{-0.12}$, and $-1.34^{+0.20}_{-0.16}$, respectively. 
Therefore, in the light of axisymmetric mass models combined with currently available data set, these dSphs suggest the presence of NFW-like or more strongly cusped dark halos.

\subsection{The impact of a sample selection and observational errors on best-fit parameters}
As indicated above, we find that dSphs in the MW and M31 have non-spherical dark halos, where large and diffuse dSphs are characterized by low dark matter densities.
However, it is unclear how the ways of data sampling (such as data volume and distribution of member stars) as well as observational errors (such as in the half-light radius, projected axial ratio of luminous parts, and distance from an observer) affect the estimation of the halo parameters.
Therefore, in what follows, we take into account these effects in the likelihood analysis and investigate their impact in the best-fit dark halo parameters.

\begin{figure}[t!!!]
\figurenum{4}
  \begin{center}
   \includegraphics[width=80mm]{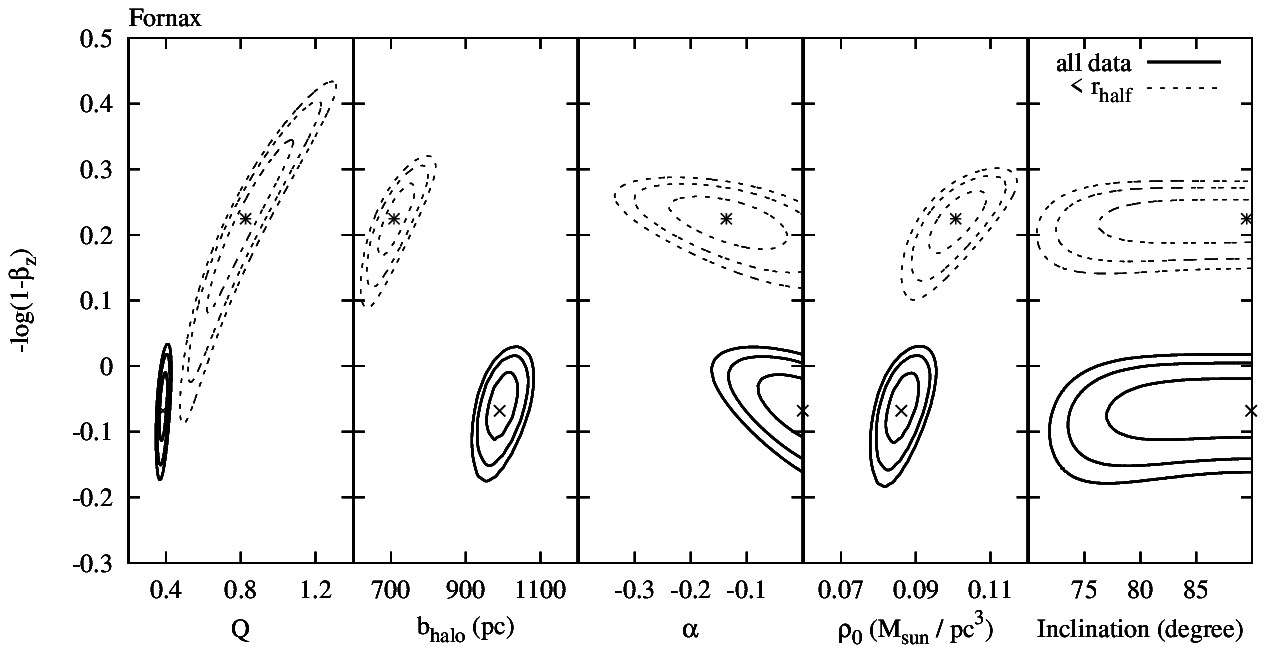}
  \end{center}
  \begin{center}
   \includegraphics[width=80mm]{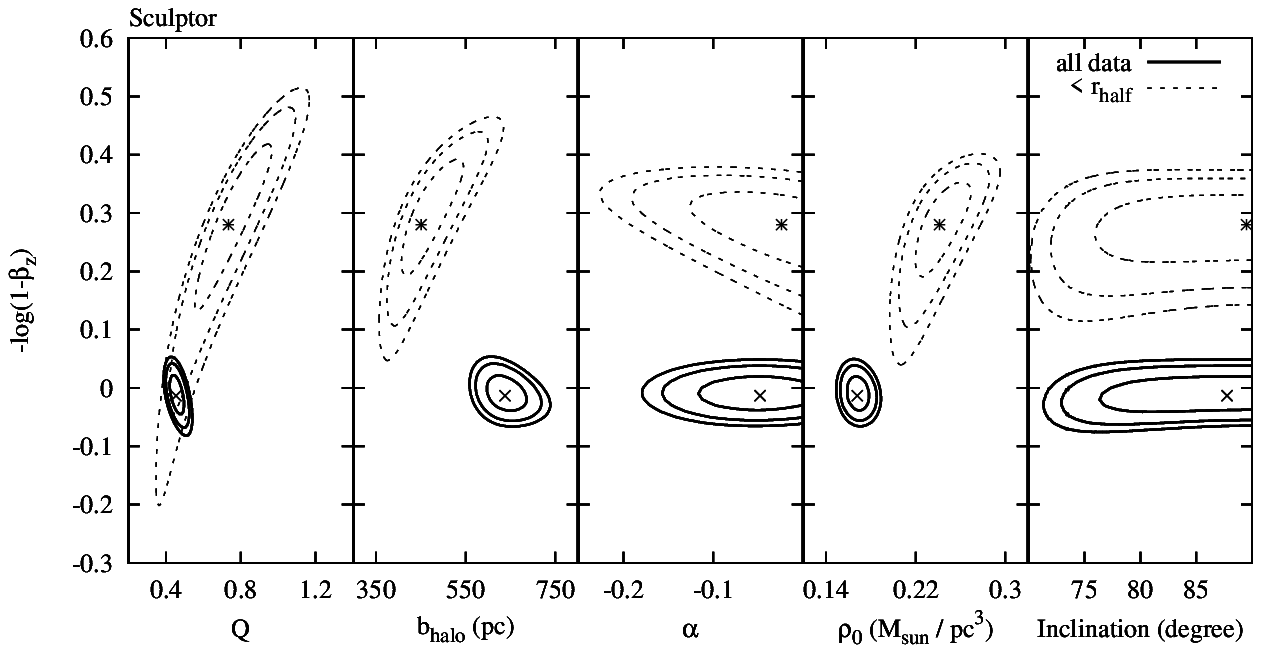}
  \end{center}
 \caption{Comparison of likelihood contours calculated from different data selections for Fornax (upper panel) and Sculptor (lower panel). Contours drawn with solid and dashed lines show 68\% $(1\sigma)$, 95\%, and 99\% confidence levels calculated from full data sample and data only within their half-light radius, respectively.}
  \label{fig:fig4}
 \end{figure}

First, in order to inspect the effects of data volume and distribution of member stars, we derive line-of-sight velocity dispersions using the data only within half-light radii for Fornax and Sculptor dSphs, and then run MCMC analysis.
Since these galaxies have the largest number of data sets and their member stars are widely distributed beyond their half-light radii, these are suited to assess the effects of the limited data volume and distribution of member stars.
As shown in Table~3 and Figure~\ref{fig:fig4}, we find that the best-fit halo parameters using the data only within the half-light radii are substantially different from those using the full data sample.
It is also found that using the data only within the half-light radii, the degeneracy between $Q$ and $\beta_z$ has emerged significantly compared to using the full data sample. Therefore it is difficult to determine these parameters due to the lack of data sample in the outer regions.
This is because, as described in the appendix, the impacts of $\beta_z$ on the line-of-sight velocity dispersion along the major axis appear prominently at outer parts. Therefore, if a sufficiently large number of sample stars is available in the outer region of a dSph, we can obtain tight limits on $\beta_z$ with small uncertainties.
Consequently, the best-fit parameters for the dSphs in which spectroscopic information are incomplete in their outer region (e.g., Sextans, And~I, III, V, and  VII) are subject to the effects of data deficiency, and thus we suggest that in order to derive more confidential dark halo structure in these dSphs, we require the observational data over much larger areas.  

\begin{figure}[t!!!]
\figurenum{5}
  \begin{center}
   \includegraphics[width=80mm]{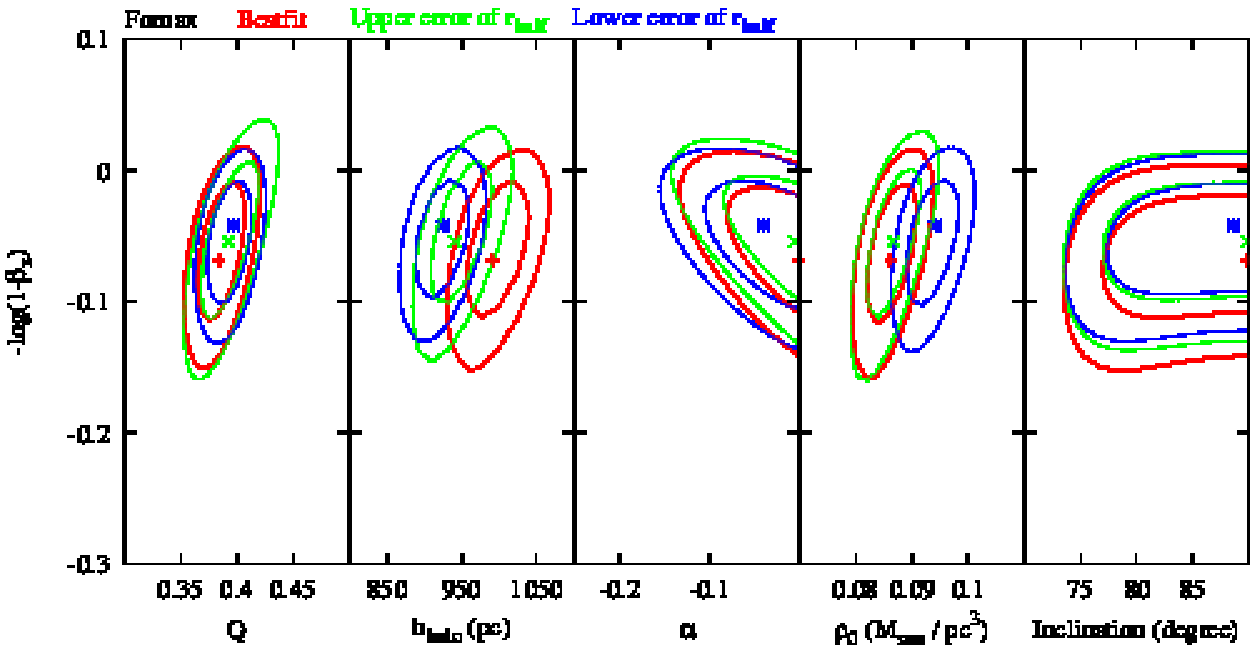}
  \end{center}
  \begin{center}
   \includegraphics[width=80mm]{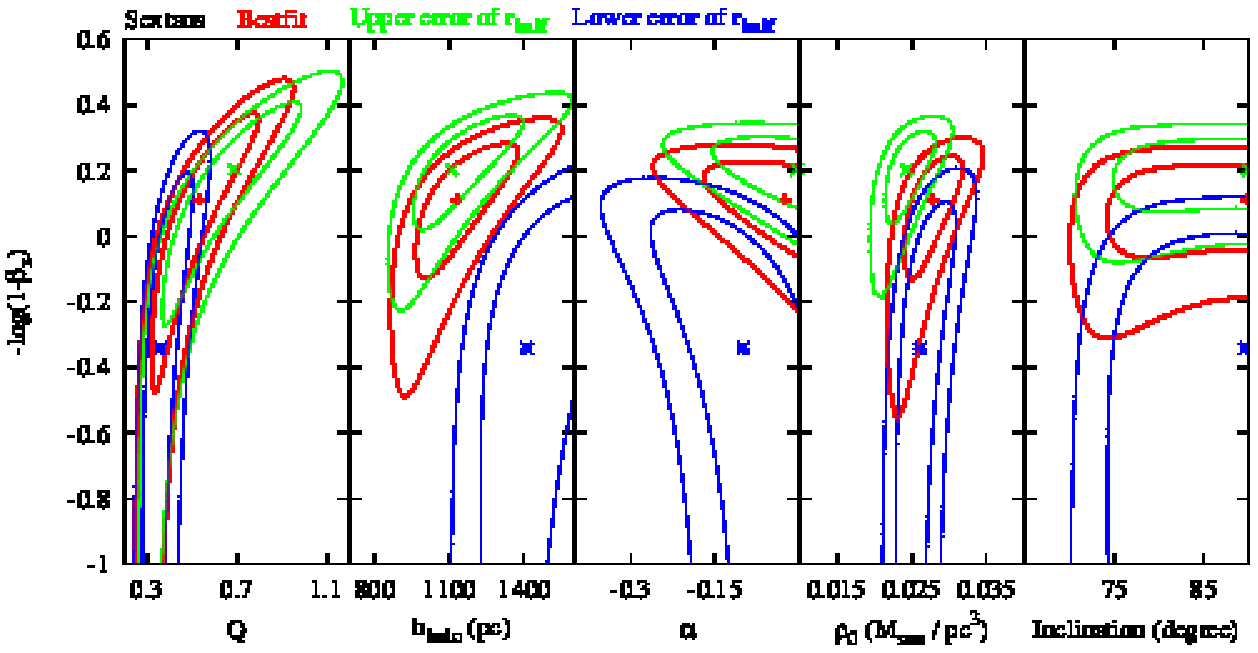}
  \end{center}
 \caption{Comparison with likelihood contours considering measurement errors in half-light radius for Foranx (upper panel) and Sextans (lower panel). Red, green, and blue contours show 68\% $(1\sigma)$ and 95\% confidence levels for the best-fit case (as shown in Figure~\ref{fig:fig2}), the cases considering an upper and lower limit in the error of $r_{\rm half}$, respectively.}
  \label{fig:fig5}
 \end{figure}

\begin{figure}[t!!!]
\figurenum{6}
  \begin{center}
   \includegraphics[width=80mm]{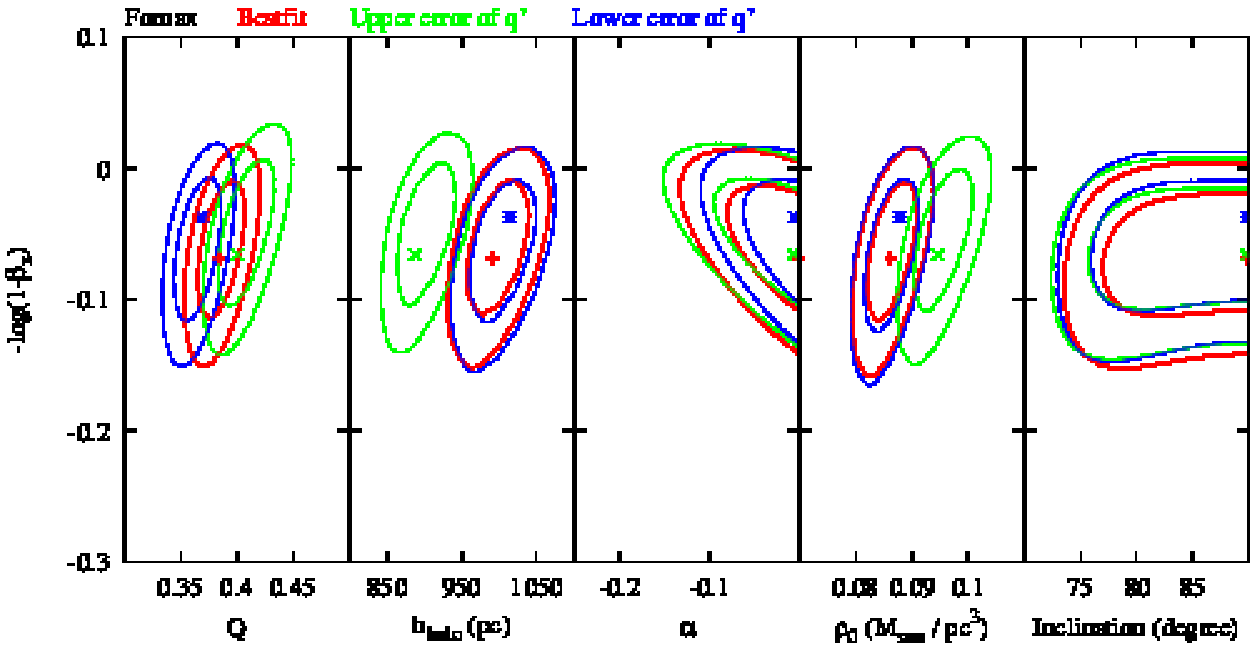}
  \end{center}
  \begin{center}
   \includegraphics[width=80mm]{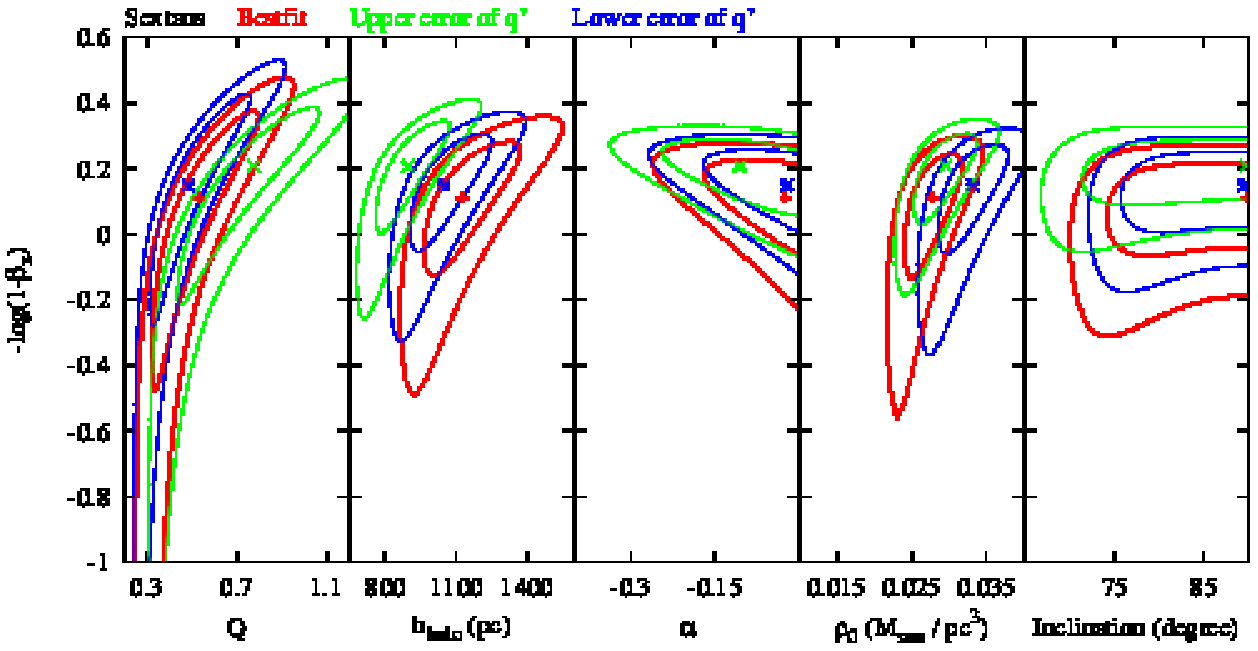}
  \end{center}
 \caption{Same as Figure~\ref{fig:fig5}, but for considering measurement errors in the projected axial ratio of the luminous component.}
  \label{fig:fig6}
 \end{figure}

\begin{figure}[t!!!]
\figurenum{7}
  \begin{center}
   \includegraphics[width=80mm]{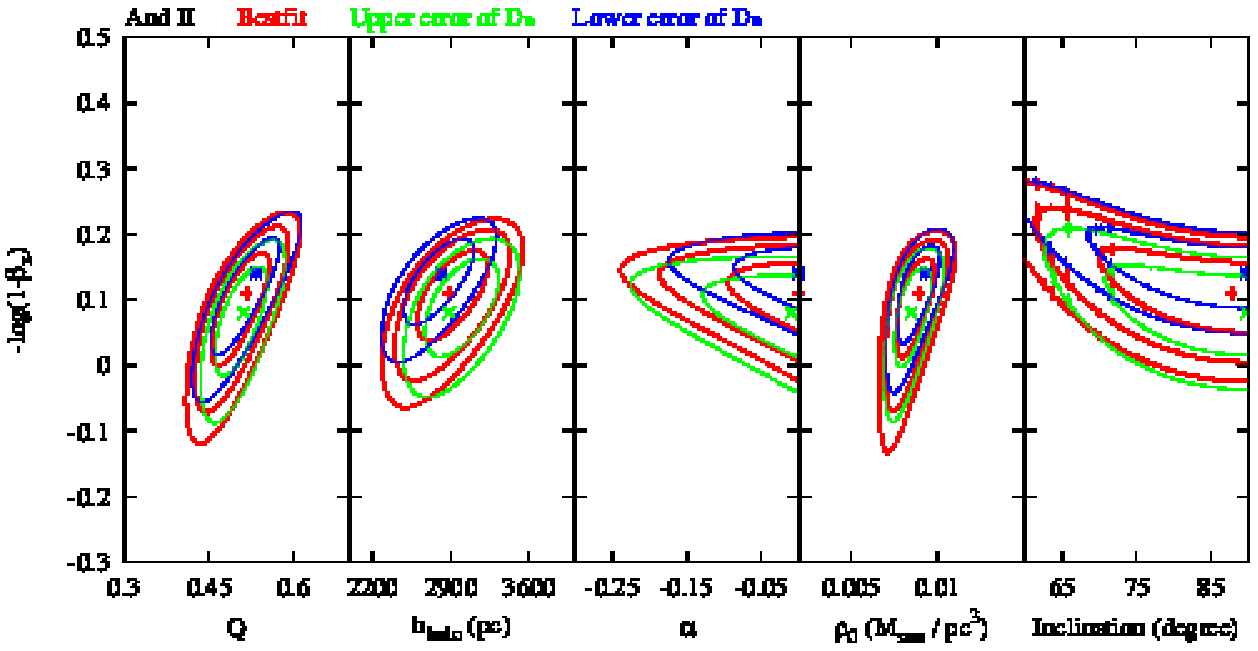}
  \end{center}
  \begin{center}
   \includegraphics[width=80mm]{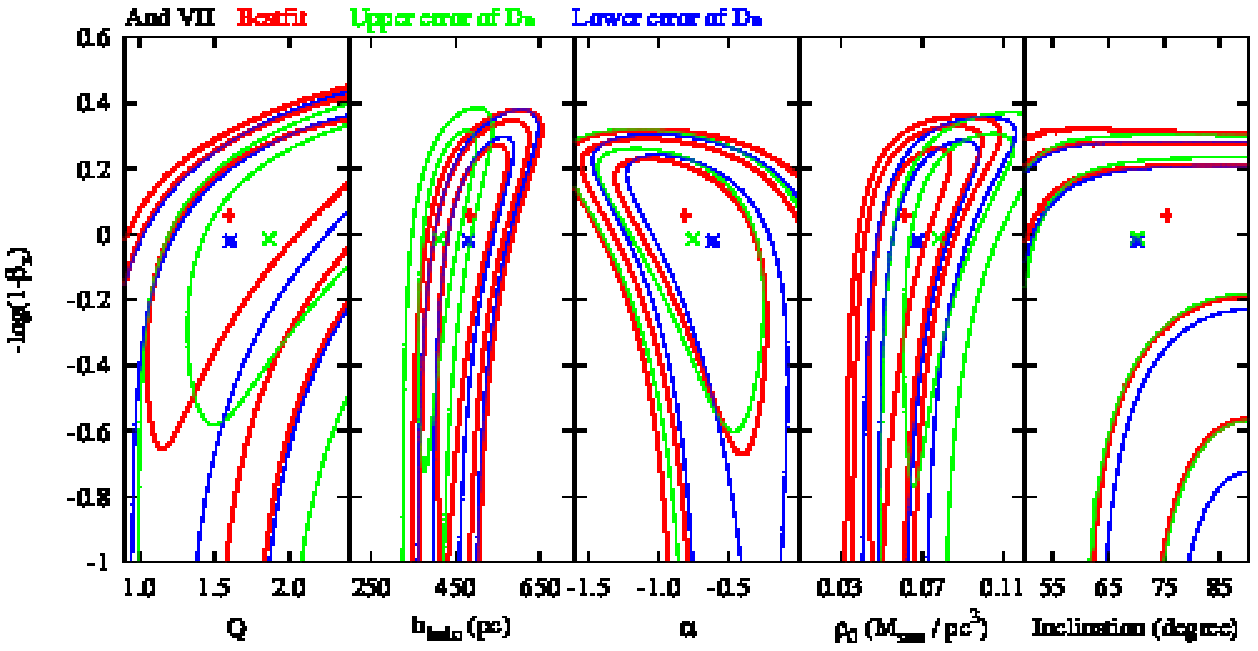}
  \end{center}
 \caption{Same as Figure~\ref{fig:fig5}, but considering measurement errors in the distance from the Sun for And~II (upper panel) and And~VII (lower panel).}
  \label{fig:fig7}
 \end{figure}

Second, we repeat the MCMC analysis to obtain the best-fit parameters considering uncertainties in the half-light radius, $r_{\rm half}$, and projected axial ratio, $q^{\prime}$, of the stellar system, because what stellar density profile is adopted affects the best-fit parameters of a dark halo through Jeans equations~\citep*{Evaetal2009}.
To investigate this, we estimate best-fit parameters for Fornax and Sextans by adopting the upper and lower observational limits of $r_{\rm half}$ and $q^{\prime}$ within their uncertainties.
These dSphs have similar $r_{\rm half}$ and $q^{\prime}$ but their errors in Sextans are larger than those in Fornax as shown in Table~1.
Table~4 and Figure~\ref{fig:fig5} show the effects of the errors in $r_{\rm half}$ on dark halo parameters.
For Fornax, with a smaller standard error in $r_{\rm half}$, fitting results are hardly affected by this error, while a relatively large error in $r_{\rm half}$ for Sextans affects the best-fit $Q$ and $\beta_z$.
On the other hand, the effects of observational errors in $q^{\prime}$ are weak compared with those of $r_{\rm half}$ as shown in Figure~\ref{fig:fig6}. 
We note that for the $b_{\rm halo}-\beta_z$ and $\rho_0-\beta_z$ relations in Fornax, each contour deviates rather largely because of the degeneracy between $\rho_0$ and $b_{\rm halo}$. 
Likewise, we investigate the impact of measurement errors in the distance from the Sun using the data of And~II and VII dSphs which show large uncertainties in the distance estimation compared with MW ones (see Table~1).
As shown in Figure~\ref{fig:fig7}, we find that the distance error as reported for these dSphs does not affect the results of maximum likelihood analysis. 

Accordingly, the constraints on dark halo structures in dSphs are affected largely by the lack of kinematic sample and distribution of member stars rather than uncertainties in photometric data.
Thus, to set robust constraints on dark halo structures in dSphs, we require deep photometric data to assemble many sample stars down to faint magnitudes and spectroscopic data over large areas out to the tidal radii of dSphs.

\section{DISCUSSION AND CONCLUDING REMARKS}
\subsection{Comparison with $\Lambda$CDM subhalos}
Recently, \citet[][hereafter VC14]{Veretal2014}~investigated the distribution for axial ratios of triaxial subhalos obtained from the Aquarius simulations.
They found that axial ratios of subhalos with maximum circular velocity in the range $8<V_{\rm max}<200$ km~s$^{-1}$ are little different between isolated subhalos and those associated with host galaxies.
Moreover, these subhalos can be approximated as oblate axisymmetric objects; the averages of intermediate-to-major axial ratios, $\langle b/a\rangle$, and minor-to-major axial ratios, $\langle c/a\rangle$,  (supposing $a\geq b\geq c$) at $r\sim1$ kpc are $\langle b/a\rangle\sim0.75$ and  $\langle c/a\rangle\sim0.60$, respectively.
They also analyzed the velocity structure in the cylindrical radial and vertical directions, i.e., $\sigma_R$ and $\sigma_z$, respectively, estimated anisotropy $\beta_z= 1-\sigma^2_z/\sigma^2_R$, and computed this as a function of distance along minor and major axes.
Their analysis shows that along the minor axis $\sigma_z$ is almost the same as $\sigma_R$, that is, $\beta_z\sim0$, whereas along the major axis $\beta_z>0$ near the halo center and $\beta_z<0$ at the outer parts.
This systematic trend is, however, weak with large scatters among the individual subhalos, thus, the radial dependence of $\beta_z$ may be regarded relatively weak.


Although these simulation results may support the validity of our axisymmetric models for the dark halos of dSphs, the axial ratios of dark halos in both MW and M31 dSphs are systematically smaller, and thus more flattened, than those of $\Lambda$CDM subhalos.
We note that for this comparison with VC14's simulations, in the case of prolate halos ($Q>1$), where the minor axis of a dark halo is aligned with the major axis of a stellar distribution, we use $Q^{\prime}=1/Q$ instead of $Q$ following VC14, so $Q^{\prime}=0.41$ for And~I and 0.21 for And~V.
As shown in Figure~7 in VC14, there are no subhalos with axial ratios less than 0.6, which accords with the interpretation that less-massive halos tend to be more relaxed and therefore more spherical than massive halos. 
\citet{Kuhetal2007}~and~\citet{Schetal2012} also have concluded from their $N$-body simulations that less-massive dark halos have mildly triaxial shapes. 
In contrast, our present analysis for the dark halos of dSphs suggests that $Q$ is much smaller than the prediction of these simulations. 
Thus, there exists a mismatch in the shapes of subhalos predicted from $\Lambda$CDM theory.

One of the possible mechanisms to alleviate this discrepancy may be tidal effects from a host dark halo.
Recent $N$-body simulations show that subhalos become significantly elongated at pericenter passage~\citep[e.g.,][]{Kuhetal2007} because of undergoing strong tidal force from deep potential of its host dark halo.
Moreover, \citet{Baretal2015} predicted that dSphs with lower $M/L$s have more spherical dark halos due to tidal stripping, whilst more dark matter dominated systems are more triaxial. In light of these predictions, it is interesting to remark that the dark halos of Fornax, Sculptor, Leo~I, and Leo~II with relatively low $M/L$s tend to be less flattened than Carina, Sextans, and Draco with high $M/L$s.
To quantify this property, we calculate the average of dark halo axial ratios, $<Q>$, for these low and high  M/L dSphs, respectively, and obtain $<Q>_{\rm low}=0.65\pm0.08$ and $<Q>_{\rm high}=0.42\pm0.04$. Therefore, the trend for the shapes of subhalos predicted by \citet{Baretal2015} are roughly consistent with those for the shapes of dark halos associated with MW dSphs.

\begin{figure}
\figurenum{8}
\begin{center}
\includegraphics[width=80mm]{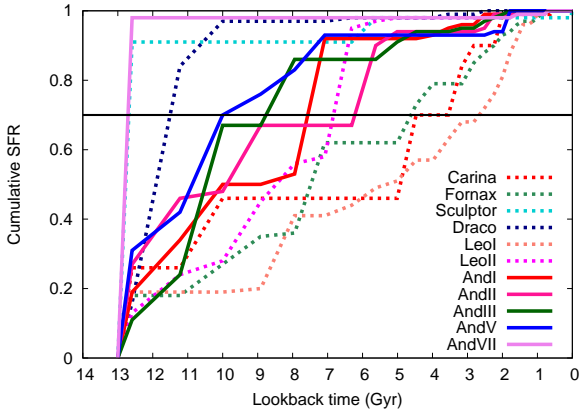}
\end{center}
\caption{Cumulative star formation history of dwarf satellites in the MW and M31 taken from~\citet{Weietal2014}.}
\label{fig:fig8} 
\end{figure}

\begin{figure}
\figurenum{9}
\begin{center}
\includegraphics[width=85mm]{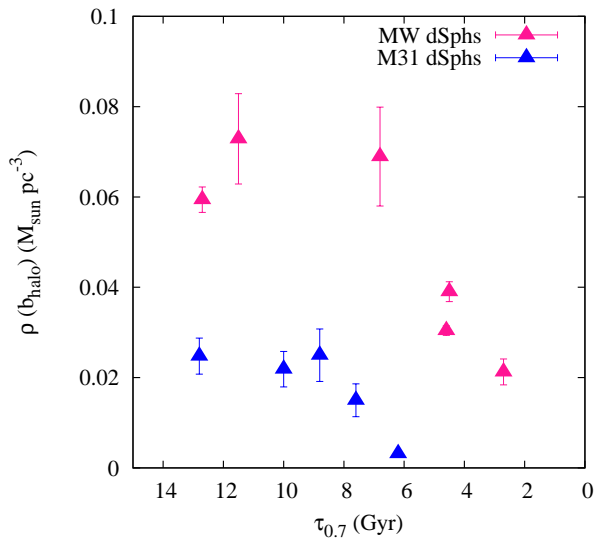}
\end{center}
\caption{Relation between $\rho (b_{\rm halo})$ and the lookback time achieving $70\%$ of current stellar masses.}
\label{fig:fig9} 
\end{figure}

However, the axial ratio of these simulated subhalos at the closest pericentric distance is never smaller than $0.5$, and thus we should consider additional mechanisms to make a dark halo more flattened.
We suggest that baryons may potentially invoke reproduction of very flattened dark halos.
First, owing to adiabatic contraction during disk formation \citep[e.g.,][]{Dutetal2007}, a dark halo with a massive disk galaxy can have a deep and steep gravitational potential in its central part compared with a pure dark halo without baryons.  
Then, when subhalos pass through this deep potential of the host halo, the shape of subhalos may be more flattened by strong tidal distortion than the case without considering baryonic effects in the host halo.
Second, as envisaged by \citet{Bryetal2013}, while cooled baryons tend to round a dark halo~\citep{Kazetal2004}, their removal due to strong feedback in the central parts of dSphs, as seen in recent simulation studies~\citep[e.g.,][]{Govetal2010,Teyetal2013,DiCetal2014a,DiCetal2014b,Madetal2014}, may prevent dark halos of such dSphs from being rounder, i.e. their shapes remain flattened.
It is, however, still unclear how the combination of these baryonic effects indeed modifies the shapes of dark subhalos. To resolve these issues, while more observational data are to be assembled for more robust determination of dark halo parameters, further simulation studies fully taking into account baryonic effects are worth exploring to get important insights into the issue of dark halo shapes addressed here.

\subsection{The relation between dark-halo structure and SFH}
As mentioned in the previous subsection, recent numerical simulations imply that internal dark matter structure in the low-mass halos can be altered by stellar feedbacks associated with star formation activity. Therefore, it is natural to expect that dark halo properties depend on the SFH of the stellar components, and we thus investigate whether this dependence indeed exists by comparing it with the observed SFH of dSphs.
For this purpose, we adopt the recent work by~\citet{Weietal2014}, which derive the SFH of dwarf galaxies in the Local Group based on the analysis of their color-magnitude diagram taken from deep imaging of the {\it Hubble Space Telescope}.
Figure~\ref{fig:fig8} displays the cumulative SFH of dwarf satellites in the MW (except for Sextans) and M31 taken from their work, indicating various formation histories such as rapid and consecutive star formation.
To facilitate the comparison with the present work, we estimate the lookback time at achieving $70\%$ of the current stellar mass of these dSphs, $\tau_{0.7}$ (as indicated as a black horizontal line in Figure~\ref{fig:fig8} and this value in each dSph is shown in Table~1); $\tau_{0.7}$ may characterize the duration and efficiency of star formation in dSphs.
This is because if we choose $50\%$ instead of $70\%$ of the current stellar mass as an indicator, we cannot completely trace the all activity phases of star formation that would be capable of changing dark-halo structures. For instance, as is shown in Figure~\ref{fig:fig8}, Fornax, Leo~I, and And~II have the periods of active star formation after the stellar masses exceed a half of these current ones.
We compare between $\tau_{0.7}$ and dark halo parameters obtained from our analysis.
Figure~\ref{fig:fig9} shows the relation between $\rho (b_{\rm halo})$ and $\tau_{0.7}$.
We consider MW and M31 dSphs separately because they are located in the different host halos and there is actually a systematic difference in quenching times of their star formations due to environmental effects as suggested by \citet{Weietal2015}. It is found that dSphs showing relatively consecutive SFH, namely lower values of $\tau_{0.7}$ (e.g., Leo~I, Fornax, and And~II) have a low dark matter density within a central region, whereas those with rapid SFH (e.g., Sculptor, Draco, and And~VII) have a dense and concentrated dark halo.
We interpret that the central density of a dark halo may be gradually decreased associated with gaseous outflow driven by effects of stellar feedbacks.

We also inspect other possible relations between $\tau_{0.7}$ and dark halo parameters such as an axial ratio, $Q$, and a central density slope, $\alpha$, but there are no remarkable relationships within uncertainties of these parameters. 
This is because, as mentioned above, the shapes of dark halos would be changed by not only stellar feedbacks, but also tidal effects from a host galaxy and dark halo.
On the other hand, for central density slopes of a dark halo, \citet{NB2015}~and \citet{DP2015}~proposed the possible mechanism of the core formation that the main process of transforming cusps to cores would be dynamical friction of gas clumps. Moreover, it may give only a small impact on $\rho(b_{\rm halo})$, because this transformation occurs in the innermost regions of a dark halo.
According to their calculation, this process may occur prior to star formation and subsequent feedback processes, implying that the relation between $\alpha$ and SFH may be made hazy, in agreement with the observed properties.
Finally, we note that we find no explicit relation between dark halo parameters and averaged metallicity, $<$[Fe/H]$>$, which is taken from~\citet{Kiretal2013}.
This may be due to the fact that $<$[Fe/H]$>$ in the stars reflect several processes such as a gas outflow as well as SFH, so the direct relation with background dark-halo properties may be weakened.

\begin{figure}
\figurenum{10}
  \begin{center}
   \includegraphics[width=85mm]{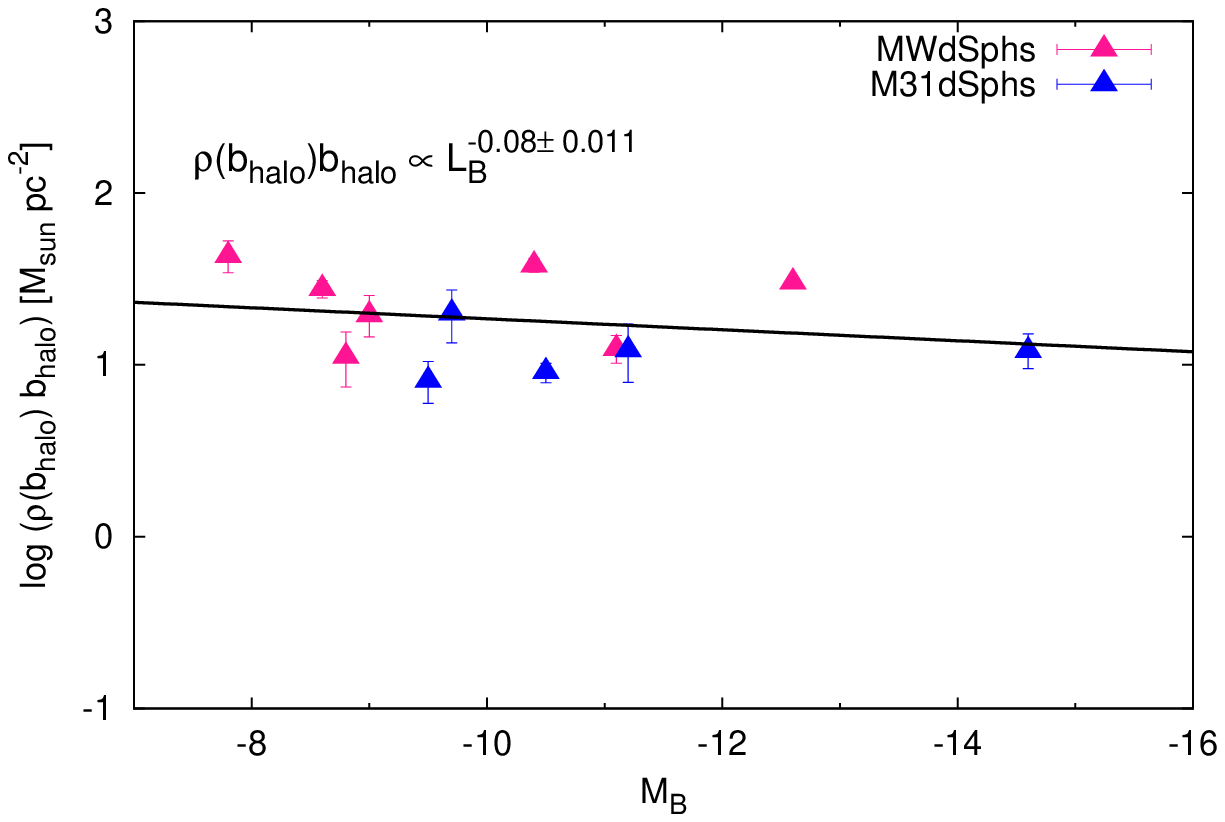}
  \end{center}
  \begin{center}
   \includegraphics[width=85mm]{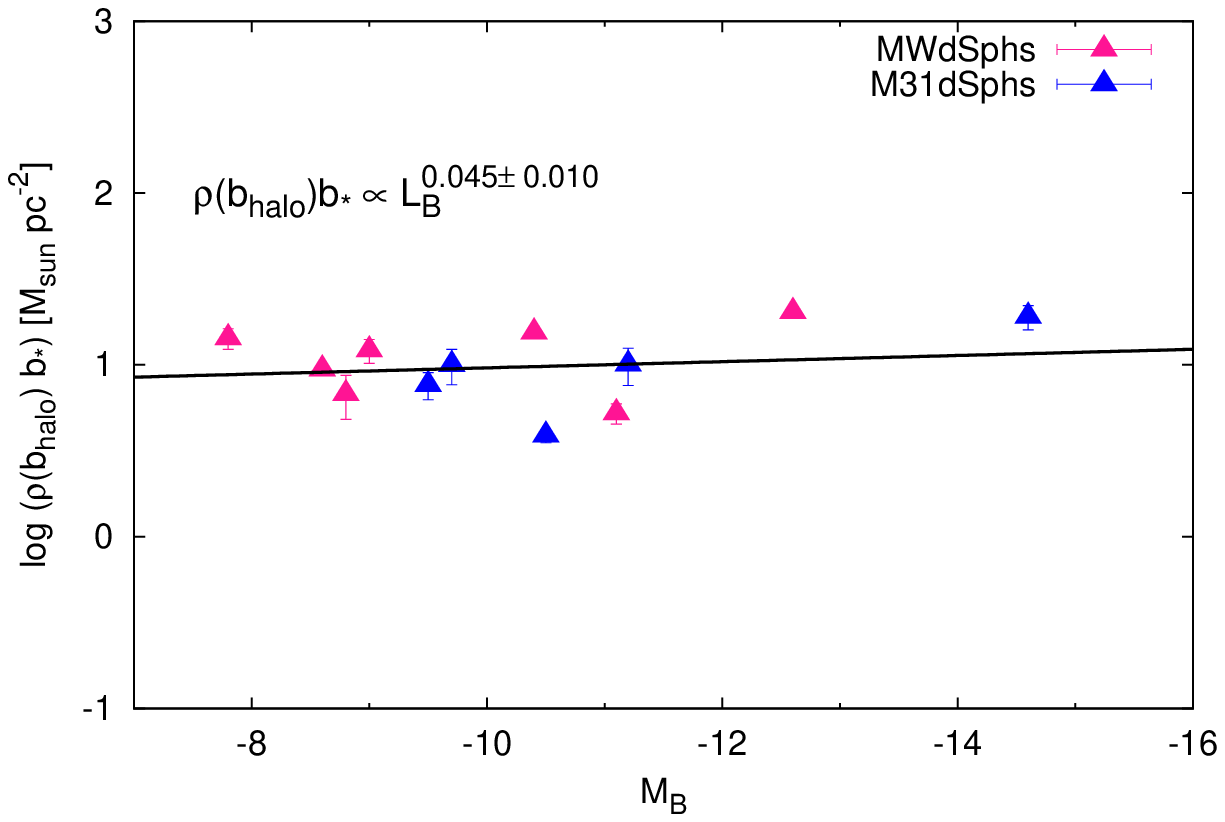}
  \end{center}
 \caption{Product of $\rho(b_{\rm halo})b_{\ast}$ (upper panel) and  $\rho(b_{\rm halo})b_{\rm halo}$ (lower panel) as a function of B-band absolute magnitude. The solid lines and in each panel denote our best-fit line to the data using least squares method.}
  \label{fig:fig10}
 \end{figure}

Thus, although both the available data and mass modeling for dSphs are yet limited, we suggest the possible link between the density of their dark halos and SFH of the dSphs. 
In order to explore this relation in more detail, we require not only further refined modeling such as considering triaxial structures of dark halos, but also assembling more photometric and kinematic data of the dSphs in their larger parts.

\subsection{Constant dark matter surface density within a radius of maximum circular velocity} 

Some previous studies have noted an intriguing general property of dark halos in a various kinds of galaxies.
For the studies of~\citet{KF2004,KF2014}, using 55 spiral galaxy rotation curves and the line-of-sight velocity dispersions of a few dSphs, they evaluated the central density $\rho_0$ and core radius $r_c$ of dark halo assuming the non-singular isothermal sphere models, and they first found that $\rho_0r_c$ is almost independent of galaxy luminosity, $\rho_0r_c\propto L_{B}^{0.058\pm0.067}$.
Similarly, \citet{Genetal2009}, \citet{Donetal2009},~and \citet{Saletal2012}~also found and confirmed, respectively,  the constant dark matter surface density, despite of assuming any cored dark matter density profiles such as Burkert and pseudo-isothermal models.
These works have provided us considerable insight for the universal properties of dark-halo structure, and thus we investigate whether this constancy exists in our estimated dark halos in dSphs.
Figure~\ref{fig:fig10} displays the relation between dark matter surface densities and luminosities of each dSph.
The upper panel shows the product of dark matter densities at its scale length, $\rho(b_{\rm halo})$, and half-light radius of a stellar component, $\rho(b_{\rm halo})b_{\ast}$, whilst the lower panel is $\rho(b_{\rm halo})b_{\rm halo}$.
In order to investigate the constancy of above two relations, we employ a least squares fitting method to determine the slope of these surface densities as a function of luminosity, and we find that the slopes of those are $\rho(b_{\rm halo})b_{\ast}\propto L_{B}^{0.045\pm0.010}$,  and $\rho(b_{\rm halo})b_{\rm halo}\propto L_{B}^{-0.08\pm0.011}$, respectively. 
It is clear that $\rho(b_{\rm halo})b_{\ast}$ and $\rho(b_{\rm halo})b_{\rm halo}$ are nearly constant values with respect to the luminosity of these galaxies.
In comparison with above previous results, our results have a similar constancy, even though assumed dark halos have cusped profiles.

\begin{figure*}
\figurenum{11}
\plotone{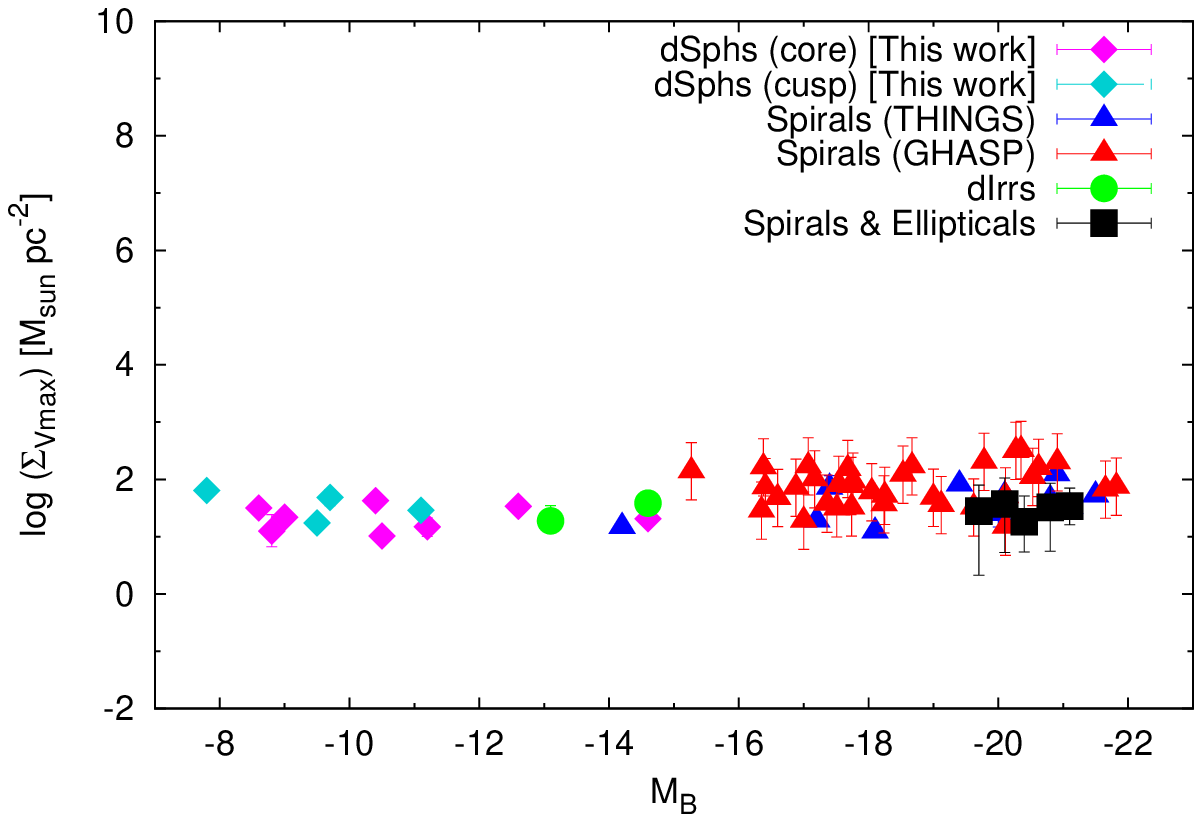}	
\caption{Mean surface dark halo density, $\Sigma_{V_{\rm max}}$ as a function of B-band absolute magnitude.
The rhombic points denote the dSphs, which have cored (magenta) and cusped (cyan) dark halos, based on this work. 
On the other hand, the other sample data are the original sample of~\citet{Donetal2009}.
Blue and red triangles are the nearby spirals in THINGS~\citep{deBetal2008} and GHASP~\citep{Spaetal2008} data, respectively.
Green circles and black squares are dwarf irregular galaxies~\citep{Genetal2005,Genetal2007}, and spiral and elliptical galaxies investigating by weak lensing (Hoekstra et al. 2005), respectively. 
The dark halo density distribution of these sample galaxies are assumed by any cored profiles such as Burkert and pseudo-isothermal profile.}
\label{fig:fig11} 
\end{figure*}

However, because of the difference of assumed dark halo density profiles among this comparison, the definitions of dark matter surface densities of each study should also be different.
Thus, in order to inspect more universal dark halo properties without the ambiguously of the assumed dark halo density profile, we introduce the new definition for dark halo surface density, namely, the mean dark halo surface density within a radius of maximum circular velocity, $\Sigma_{V_{\rm max}}$.
Since this surface density was already defined in our recent paper~\citep[][hereafter HC15]{HC2015}, we describe only the important properties of this. 
This surface density is directly correlated with the central dark matter surface density, $\Sigma_{V_{\rm max}} \propto \rho_{\ast}r_{\ast}$, where $\rho_{\ast}$ and $r_{\ast}$ are central density and scale length of arbitrary dark halo profiles, respectively.
In HC15, we calculated $\Sigma_{V_{\rm max}}$ using the data of dark halo profiles based on the HI gas rotation curve of late- and early-type spirals with pseudo-isothermal dark halos~\citep{deBetal2008,Spaetal2008}, dwarf irregulars with Burkert dark halos~\citep{Genetal2005,Genetal2007}, the galaxy--galaxy weak lensing from spiral and elliptical galaxies with Burkert profiles~(analyzed by \citealt{Donetal2009}, data from \citealt{Hoeetal2005}) and the dSphs in this work. Then we found that $\Sigma_{V_{\rm max}}$ does not depend on the maximum circular velocity corresponding with dark-halo mass of these galaxies.
Figure~\ref{fig:fig11} shows the $\Sigma_{\rm M, half}-M_{B}$ relation for same sample in HC15 with luminosity over almost 14 orders of magnitude.
As shown in this figure, we confirm that this dark-halo surface density has also constancy against galaxy luminosity without concerning the assumption of dark halo profiles.

\subsection{Concluding Remarks} 
In this paper, we revisit the dynamical analysis of non-spherical dark halos for the dSphs in  the MW and M31 based on revised axisymmetric mass models.
In contrast to HC12, which assumed $\overline{v^2_z}=\overline{v^2_R}$ for simplicity; we relax this constraint and set a non-zero velocity anisotropy, $\beta_z=1-\overline{v^2_z}/\overline{v^2_R}$, as this anisotropy can be degenerate with the effect of $Q$.
Based on the application of our models to seven MW dSphs and five M31 dSphs, we find that the best-fitting cases for most of the dSphs yield not spherical, but flattened halos even considering $\beta_z$.
It is worth noting that the best-fit parameters, especially for the shapes of dark halos and velocity anisotropy, are susceptible to both the availability of kinematic data in the outer regions of the system as well as the effect of the small number of sample stars.
Thus, to estimate more plausible dark-halo structure in the dSphs, we require the photometric and kinematic data over much larger areas.
We have found from the revised analysis and currently available data that the derived shapes of dark halos in dSphs remain systematically more flattened than those of dark matter subhalos calculated from $\Lambda$CDM-based $N$-body simulations, as suggested in HC12.
This mismatch needs to be solved by a theory including baryon components and the associated feedback to dark halos as well as by further observational limits in larger areas of dSphs.
It was also found that more diffuse dark halos may have undergone consecutive SFH, as characterized by low $\tau_{0.7}$, implying that the formation process of dSphs is affected and thus imprinted in the structure of their dark halos.
In the near future, planned surveys of the MW and M31 dSphs using Hyper Suprime-Cam and Prime Focus Spectrograph attached on the Subaru Telescope~\citep{Taketal2014} will enable us to hunt a number of faint stars in the outer parts of dSphs and measure their kinematic data and metallicities, thereby allowing us to obtain severer limits on the dark halo distribution and characterize the dynamical evolution of dSphs.

\acknowledgments
We are grateful to the referee for her/his careful reading of our paper and thoughtful comments.
We would like to give special thanks to Nhung Ho and Puragra GuhaThakurta for giving us the kinematic data of And~II and for useful discussions.
We also thank Kohei Hattori, Hidetomo Honma, and Yusuke Komuro for useful discussions.
This work is supported in part by a Grants-in-Aid for Scientific Research from the Japan Society for the Promotion of Science (JSPS) (No. 26-3302 for K.H.).
This work has been supported in part by a Grant-in-Aid for Scientific Research (25287062) of the Ministry of Education, Culture, Sports, Science and Technology in Japan and by JSPS Core-to-Core Program ``International Research Network for Dark Energy.''
Finally, Kavli IPMU was established is supported by World Premier International Research Center Initiative (WPI), MEXT, Japan.



\appendix
\section{Effects of $Q$ and $\beta_z$ on line-of-sight velocity dispersion}
In this appendix, we demonstrate the impact of the non-spherical shape of a dark halo, $Q$, and the velocity anisotropy $\beta_z$ of member stars on their line-of-sight velocity dispersion profiles, motivated by the~\citet{Cap2008} paper, which showed that the variations of $Q$ and $\beta_z$ are a similar effect on $\sigma_{\rm los}$ profiles.

\begin{figure*}
 \figurenum{A.1}
 \begin{minipage}{0.5\hsize}
  \begin{center}
   \includegraphics[width=94mm]{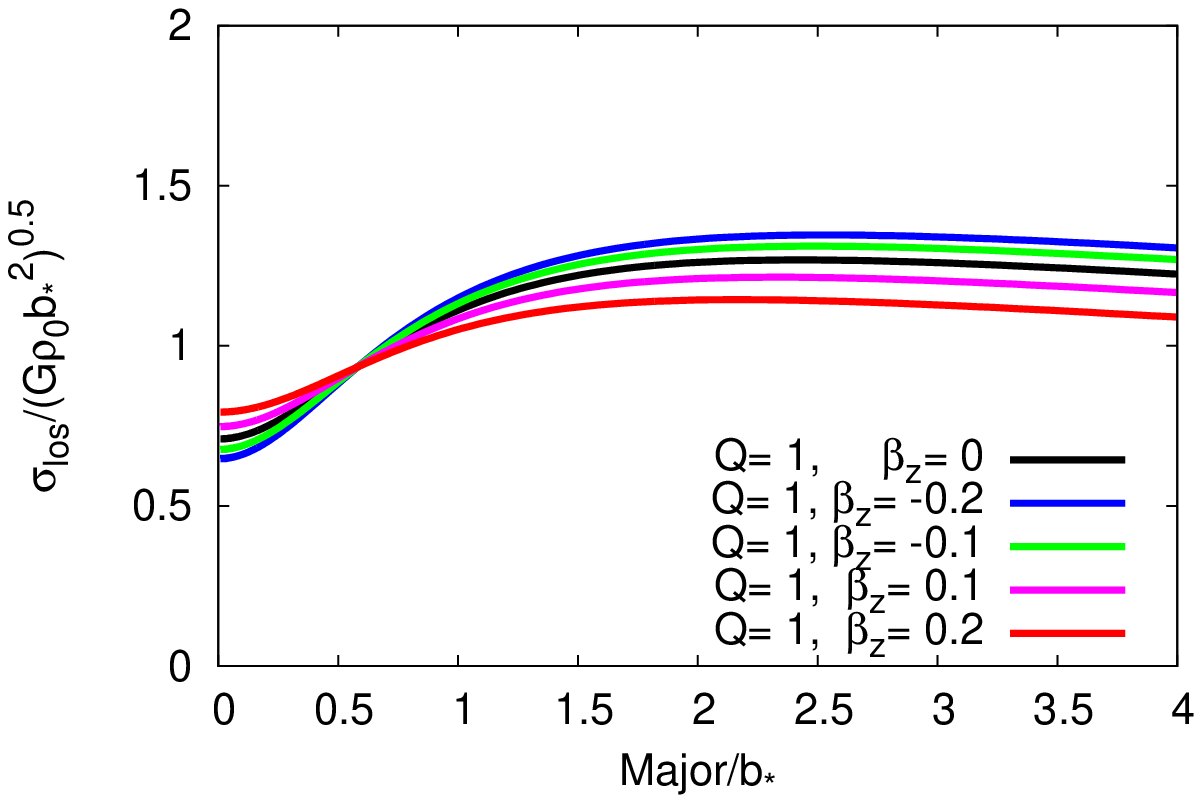}
  \end{center}
 \end{minipage}
 \begin{minipage}{0.5\hsize}
  \begin{center}
   \includegraphics[width=94mm]{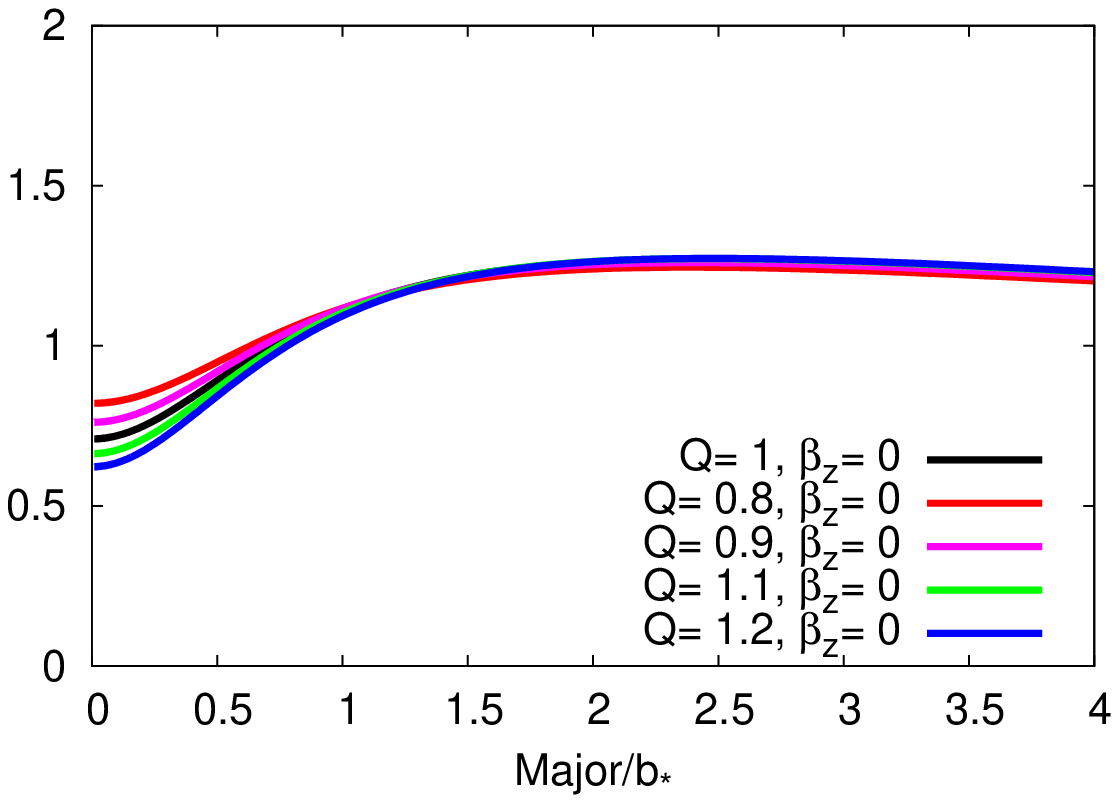}
  \end{center}
 \end{minipage}
  \begin{minipage}{0.5\hsize}
  \begin{center}
   \includegraphics[width=94mm]{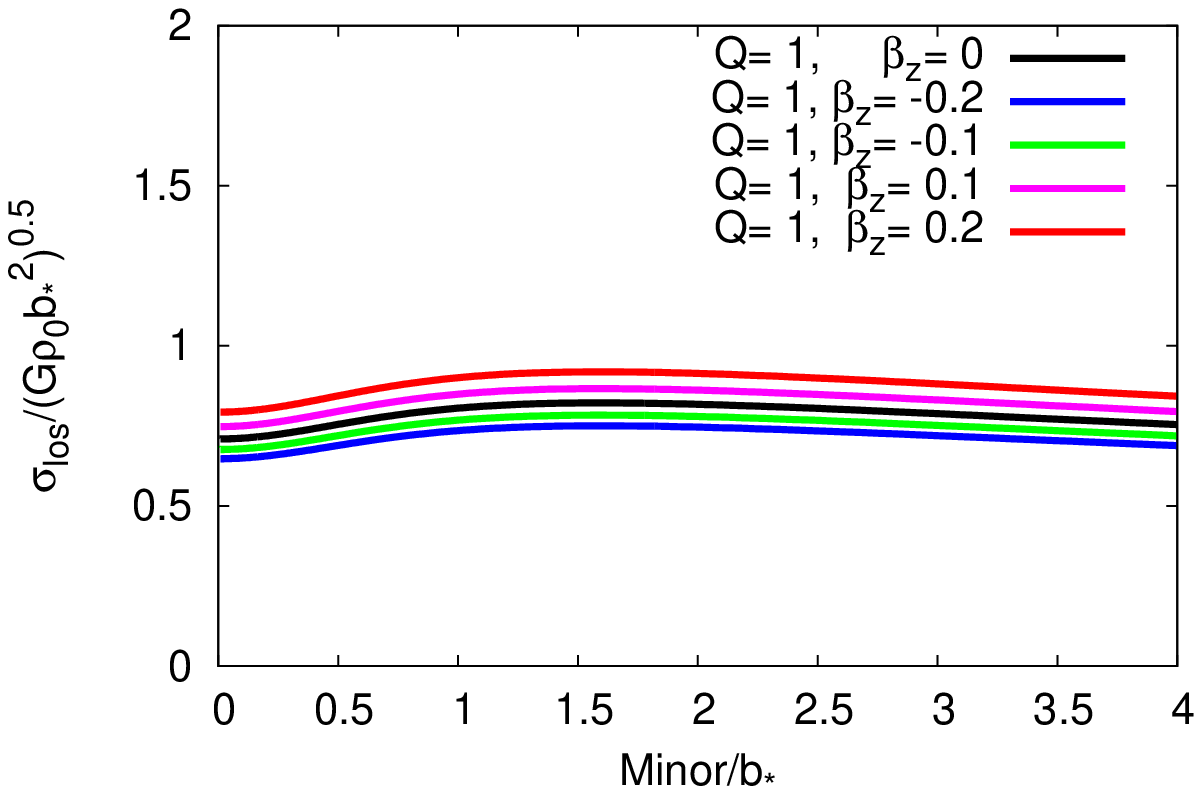}
  \end{center}
 \end{minipage}
 \begin{minipage}{0.5\hsize}
  \begin{center}
   \includegraphics[width=94mm]{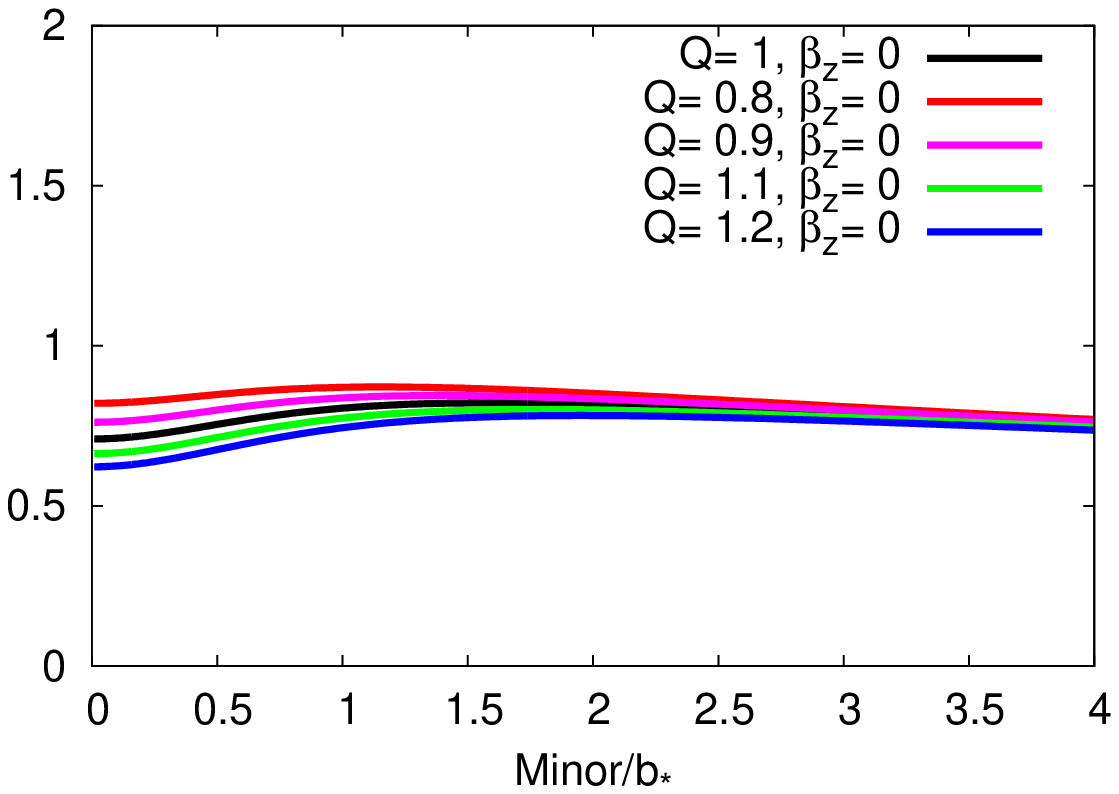}
  \end{center}
 \end{minipage}
\caption{Upper panels show normalized line-of-sight velocity dispersions, $\sigma_{\rm los}/(G\rho_{0}b^2_{\ast})^{1/2}$, along the major axis for a cored case ($\alpha=0$), whereas the bottom panels show these velocity dispersions along the minor axis for the same case. For all of these cases we suppose that the edge-on galaxy ($i=90^{\circ}$) and the ratio of $b_{\rm halo}/b_{\ast}$ is unity for the sake of demonstration.}
  \label{fig:figA1}
 \end{figure*}

Figure~\ref{fig:figA1} indicates normalized line-of-sight velocity dispersions along the major and minor axes, respectively.
We set the fixed axial ratios of stellar distribution in this test calculation, $q=0.7$, which is a typical value in MW dSphs.
First, as is shown in the upper panels of Figure~\ref{fig:figA1} denoting the $\sigma_{\rm los}$ profiles along the major axis, 
the effects of increasing $\beta_z$ (top left) and of decreasing $Q$ (top right) are resemblant in the property that these weaken the wavy feature of the $\sigma_{\rm los}$ profile caused by a non-spherical $Q$ along the major axis, as already discussed in HC12.
However, there is a crucial difference between these effects. 
The difference in $\sigma_{\rm los}$ profiles by the change of $\beta_z$ does occur at not the inner but outer parts. This is because $\overline{v^2_{\phi}}$ including a constant velocity anisotropy in Equation~(2) contributes to $\sigma_{\rm los}$ more than $\overline{v^2_{z}}$ dose in outer parts, thus the impacts of $\beta_z$ on $\sigma_{\rm los}$ emerges at only outer parts.
On the other hand, for the case of the change of $Q$, this is opposed to that of $\beta_z$. 
This is because a larger (smaller) $Q$ yields weaker (stronger) gravitational force in the $z$-direction, thereby decreasing (increasing) $\overline{v^2_z}$ in inner parts.

Second, as shown in the lower panels of Figure~\ref{fig:figA1}, there is no difference by changing $\beta_z$,
because $\sigma_{\rm los}$ along the minor axis is not contributed by $\overline{v^2_{\phi}}$, that is, we need not to consider the effects of $\beta_z$.
On the other hand, the effects of changing $Q$ are monotonous, which can be straightforwardly understood.
As mentioned above, we can consider only the impact on $\overline{v^2_{z}}$, thus a decreasing (increasing) $Q$ reduces (increases) $\overline{v^2_z}$ in inner parts.

In summary, a finite degeneracy between $Q$ and $\beta_z$ can be broken by investigating a systematic difference in the effects of these parameters on $\sigma_{\rm los}$ profiles between the minor and major axes.
This suggests that to definitely determine $Q$, an axial ratio of a dark halo, we need a sufficiently large number of kinematic data for member stars, whereby accurate $\sigma_{\rm los}$ profiles along both minor and major axes are available.

\begin{thebibliography}{}
\bibitem[Agnello \& Evans(2012)]{AE2012} Agnello, A., \& Evans, N.~W.\ 2012, \apjl, 754, LL39 
\bibitem[Allgood et al.(2006)]{Alletal2006} Allgood, B., Flores, R.~A., Primack, J.~R., et al.\ 2006, \mnras, 367, 1781 
\bibitem[Amorisco et al.(2013)]{Amoetal2013} Amorisco, N.~C., Agnello, A., \& Evans, N.~W.\ 2013, \mnras, 429, L89 
\bibitem[Barber et al.(2015)]{Baretal2015} Barber, C., Starkenburg, E., Navarro, J.~F., \& McConnachie, A.~W.\ 2015, \mnras, 447, 1112 
\bibitem[Battaglia et al.(2011)]{Batetal2011} Battaglia, G., Tolstoy, E., Helmi, A., et al.\ 2011, \mnras, 411, 1013 
\bibitem[Binney \& Tremaine(2008)]{BT2008} Binney, J., \& Tremaine, S. 2008, Galactic Dynamics (2nd ed.; Princeton, NJ: Princeton Univ. Press)  
\bibitem[Boylan-Kolchin et al.(2011)]{Boyetal2011} Boylan-Kolchin, M., Bullock, J.~S., \& Kaplinghat, M.\ 2011, \mnras, 415, L40 
\bibitem[Boylan-Kolchin et al.(2012)]{Boyetal2012} Boylan-Kolchin, M., Bullock, J.~S., \& Kaplinghat, M.\ 2012, \mnras, 422, 1203 
\bibitem[Breddels \& Helmi(2013)]{BH2013} Breddels, M.~A., \& Helmi, A.\ 2013, \aap, 558, AA35 
\bibitem[Brook \& Di Cintio(2014)]{BD2014} Brook, C.~B., \& Di Cintio, A.\ 2015, \mnras, 450, 3920  
\bibitem[Brooks \& Zolotov(2014)]{BZ2014} Brooks, A.~M., \& Zolotov, A.\ 2014, \apj, 786, 87 
\bibitem[Bryan et al.(2013)]{Bryetal2013} Bryan, S.~E., Kay, S.~T., Duffy, A.~R., et al.\ 2013, \mnras, 429, 3316 
\bibitem[Burkert(1995)]{Bur1995} Burkert, A.\ 1995, \apjl, 447, L25 
\bibitem[Cappellari(2008)]{Cap2008} Cappellari, M.\ 2008, \mnras, 390, 71 
\bibitem[de Blok(2010)]{deB2010} de Blok, W.~J.~G.\ 2010, Advances in Astronomy, 2010, 789293 
\bibitem[de Blok et al.(2001)]{deBetal2001} de Blok, W.~J.~G., McGaugh, S.~S., Bosma, A., \& Rubin, V.~C.\ 2001, \apjl, 552, L23 
\bibitem[de Blok et al.(2008)]{deBetal2008} de Blok, W.~J.~G., Walter, F., Brinks, E., et al.\ 2008, \aj, 136, 2648 
\bibitem[Del Popolo \& Pace(2015)]{DP2015} Del Popolo, A., \& Pace, F.\ 2015, arXiv:1502.01947 
\bibitem[Di Cintio et al.(2014a)]{DiCetal2014a} Di Cintio, A., Brook, C.~B., Dutton, A.~A., et al.\ 2014, \mnras, 441, 2986 
\bibitem[Di Cintio et al.(2014b)]{DiCetal2014b} Di Cintio, A., Brook, C.~B., Macci{\`o}, A.~V., et al.\ 2014, \mnras, 437, 415 
\bibitem[Diemand et al.(2008)]{Dieetal2008} Diemand, J., Kuhlen, M., Madau, P., et al.\ 2008, \nat, 454, 735 
\bibitem[Donato et al.(2009)]{Donetal2009} Donato, F., Gentile, G., Salucci, P., et al.\ 2009, \mnras, 397, 1169 
\bibitem[Dutton et al.(2007)]{Dutetal2007} Dutton, A.~A., van den Bosch, F.~C., Dekel, A., \& Courteau, S.\ 2007, \apj, 654, 27 
\bibitem[Evans, An \& Walker(2009)]{Evaetal2009} Evans, N.~W., An, J., \& Walker, M.~G.\ 2009, \mnras, 393, L50 
\bibitem[Fukushige \& Makino(1997)]{FM1997} Fukushige, T., \& Makino, J.\ 1997, \apjl, 477, L9 
\bibitem[Garrison-Kimmel et al.(2013)]{Garetal2013} Garrison-Kimmel, S., Rocha, M., Boylan-Kolchin, M., Bullock, J.~S., \& Lally, J.\ 2013, \mnras, 433, 3539 
\bibitem[Gentile et al.(2005)]{Genetal2005} Gentile, G., Burkert, A., Salucci, P., Klein, U., \& Walter, F.\ 2005, \apjl, 634, L145 
\bibitem[Gentile et al.(2009)]{Genetal2009} Gentile, G., Famaey, B., Zhao, H., \& Salucci, P.\ 2009, \nat, 461, 627 
\bibitem[Gentile et al.(2007)]{Genetal2007} Gentile, G., Salucci, P., Klein, U., \& Granato, G.~L.\ 2007, \mnras, 375, 199 
\bibitem[Gilmore et al.(2007)]{Giletal2007} Gilmore, G., Wilkinson, M.~I., Wyse, R.~F.~G., et al.\ 2007, \apj, 663, 948 
\bibitem[Governato et al.(2010)]{Govetal2010} Governato, F., Brook, C., Mayer, L., et al.\ 2010, \nat, 463, 203 
\bibitem[Governato et al.(2012)]{Govetal2012} Governato, F., Zolotov, A., Pontzen, A., et al.\ 2012, \mnras, 422, 1231 
\bibitem[Hastings(1970)]{Has1970}Hastings, W. K. 1970, Biometrika, 57, 97
\bibitem[Hayashi \& Chiba(2012)]{HC2012} Hayashi, K., \& Chiba, M.\ 2012, \apj, 755, 145 
\bibitem[Hayashi \& Chiba(2015)]{HC2015} Hayashi, K., \& Chiba, M.\ 2015, \apjl, 803, L11
\bibitem[Hayashi et al.(2004)]{Hayetal2004} Hayashi, E., Navarro, J.~F., Power, C., et al.\ 2004, \mnras, 355, 794 
\bibitem[Hayashi et al.(2007)]{Hayetal2007} Hayashi, E., Navarro, J.~F., \& Springel, V.\ 2007, \mnras, 377, 50 
\bibitem[Ho~et al.(2012)]{Hoetal2012} Ho, N., Geha, M., Munoz, R.~R., et al.\ 2012, \apj, 758, 124 
\bibitem[Hoekstra et al.(2005)]{Hoeetal2005} Hoekstra, H., Hsieh, B.~C., Yee, H.~K.~C., Lin, H., \& Gladders, M.~D.\ 2005, \apj, 635, 73 
\bibitem[Irwin \& Hatzidimitriou(1995)]{IH1995} Irwin, M., \& Hatzidimitriou, D.\ 1995, \mnras, 277, 1354 
\bibitem[Ishiyama et al.(2013)]{Ishietal2013} Ishiyama, T., Rieder, S., Makino, J., et al.\ 2013, \apj, 767, 146 
\bibitem[Jardel \& Gebhardt(2013)]{JG2013} Jardel, J.~R., \& Gebhardt, K.\ 2013, \apjl, 775, LL30 
\bibitem[Jardel et al.(2013)]{Jaretal2013} Jardel, J.~R., Gebhardt, K., Fabricius, M.~H., Drory, N., \& Williams, M.~J.\ 2013, \apj, 763, 91 
\bibitem[Jing \& Suto(2000)]{JS2000} Jing, Y.~P., \& Suto, Y.\ 2000, \apjl, 529, L69 
\bibitem[Jing \& Suto(2002)]{JS2002} Jing, Y.~P., \& Suto, Y.\ 2002, \apj, 574, 538 
\bibitem[Kazantzidis et al.(2004)]{Kazetal2004} Kazantzidis, S., Kravtsov, A.~V., Zentner, A.~R., et al.\ 2004, \apjl, 611, L73 
\bibitem[Kleyna et al.(2002)]{Kleetal2002} Kleyna, J., Wilkinson, M.~I., Evans, N.~W., Gilmore, G., \& Frayn, C.\ 2002, \mnras, 330, 792 
\bibitem[Kirby et al.(2013)]{Kiretal2013} Kirby, E.~N., Cohen, J.~G., Guhathakurta, P., et al.\ 2013, \apj, 779, 102 
\bibitem[Koch et al.(2007a)]{Kocetal2007a} Koch, A., Grebel, E.~K., Kleyna, J.~T., et al.\ 2007, \aj, 133, 270 
\bibitem[Koch et al.(2007b)]{Kocetal2007b} Koch, A., Kleyna, J.~T., Wilkinson, M.~I., et al.\ 2007, \aj, 134, 566 
\bibitem[Kormendy \& Freeman(2004)]{KF2004} Kormendy, J., \& Freeman, K.~C.\ 2004, Dark Matter in Galaxies, 220, 377 
\bibitem[Kormendy \& Freeman(2014)]{KF2014} Kormendy, J., \& Freeman, K.~C.\ 2014, arXiv:1411.2170 
\bibitem[Kuhlen et al.(2007)]{Kuhetal2007} Kuhlen, M., Diemand, J., \& Madau, P.\ 2007, \apj, 671, 1135 
\bibitem[Madau et al.(2014)]{Madetal2014} Madau, P., Shen, S., \& Governato, F.\ 2014, \apjl, 789, LL17 
\bibitem[Martin et al.(2008)]{Maretal2008} Martin, N.~F., de Jong, J.~T.~A., \& Rix, H.-W.\ 2008, \apj, 684, 1075 
\bibitem[Mateo(1998)]{Mat1998} Mateo, M.~L.\ 1998, \araa, 36, 435 
\bibitem[Mateo et al.(2008)]{Matetal2008} Mateo, M., Olszewski, E.~W., \& Walker, M.~G.\ 2008, \apj, 675, 201 
\bibitem[McConnachie(2012)]{McC2012} McConnachie, A.~W.\ 2012, \aj, 144, 4 
\bibitem[McConnachie \& Irwin(2006)]{MI2006} McConnachie, A.~W., \& Irwin, M.~J.\ 2006, \mnras, 365, 1263 
\bibitem[Metropolis et al.(1953)]{Metetal1953} Metropolis, A. W., Rosenbluth, M. N., Teller, A. H., \& Teller, E. 1953, J. Chem. Phys., 21, 1087
\bibitem[Moore(1994)]{Moo1994} Moore, B.\ 1994, \nat, 370, 629 
\bibitem[Navarro et al.(1996)]{NFW1996} Navarro, J.~F., Frenk, C.~S., \& White, S.~D.~M.\ 1996, \apj, 462, 563 
\bibitem[Navarro et al.(1997)]{NFW1997} Navarro, J.~F., Frenk, C.~S., \& White, S.~D.~M.\ 1997, \apj, 490, 493 
\bibitem[Nipoti \& Binney(2015)]{NB2015} Nipoti, C., \& Binney, J.\ 2015, \mnras, 446, 1820 
\bibitem[Ogiya \& Burkert(2015)]{OB2015} Ogiya, G., \& Burkert, A.\ 2015, \mnras, 446, 2363 
\bibitem[Ogiya \& Mori(2014)]{OM2014} Ogiya, G., \& Mori, M.\ 2014, \apj, 793, 46 
\bibitem[O{\~n}orbe et al.(2015)]{Onoetal2015} O{\~n}orbe, J., Boylan-Kolchin, M., Bullock, J.~S., et al.\ 2015, arXiv:1502.02036 
\bibitem[Pe{\~n}arrubia et al.(2012)]{Penetal2012} Pe{\~n}arrubia, J., Pontzen, A., Walker, M.~G., \& Koposov, S.~E.\ 2012, \apjl, 759, L42 
\bibitem[Plummer(1911)]{Plu1911} Plummer, H.~C.\ 1911, \mnras, 71, 460 
\bibitem[Salucci et al.(2012)]{Saletal2012} Salucci, P., Wilkinson, M.~I., Walker, M.~G., et al.\ 2012, \mnras, 420, 2034 
\bibitem[Schneider et al.(2012)]{Schetal2012} Schneider, M.~D., Frenk, C.~S., \& Cole, S.\ 2012, \jcap, 5, 030 
\bibitem[Spano et al.(2008)]{Spaetal2008} Spano, M., Marcelin, M., Amram, P., et al.\ 2008, \mnras, 383, 297 
\bibitem[Strigari et al.(2010)]{Stretal2010} Strigari, L.~E., Frenk, C.~S., \& White, S.~D.~M.\ 2010, \mnras, 408, 2364 
\bibitem[Strigari et al.(2014)]{Stretal2014} Strigari, L.~E., Frenk, C.~S., \& White, S.~D.~M.\ 2014, arXiv:1406.6079 
\bibitem[Takada et al.(2014)]{Taketal2014} Takada, M., Ellis, R.~S., Chiba, M., et al.\ 2014, \pasj, 66, R1 
\bibitem[Tempel \& Tenjes(2006)]{TT2006} Tempel, E., \& Tenjes, P.\ 2006, \mnras, 371, 1269 
\bibitem[Teyssier et al.(2013)]{Teyetal2013} Teyssier, R., Pontzen, A., Dubois, Y., \& Read, J.~I.\ 2013, \mnras, 429, 3068 
\bibitem[Tollerud et al.(2012)]{Toletal2012} Tollerud, E.~J., Beaton, R.~L., Geha, M.~C., et al.\ 2012, \apj, 752, 45 
\bibitem[van der Marel et al.(1994)]{vanetal1994} van der Marel, R.~P., Evans, N.~W., Rix, H.-W., White, S.~D.~M., \& de Zeeuw, T.\ 1994, \mnras, 271, 99 
\bibitem[Vera-Ciro et al.(2014)]{Veretal2014} Vera-Ciro, C.~A., Sales, L.~V., Helmi, A., \& Navarro, J.~F.\ 2014, \mnras, 439, 2863 
\bibitem[Walker et al.(2009a)]{Waletal2009a} Walker, M.~G., Mateo, M., \& Olszewski, E.~W.\ 2009, \aj, 137, 3100 
\bibitem[Walker et al.(2009b)]{Waletal2009b} Walker, M.~G., Mateo, M., Olszewski, E.~W., Sen, B., \& Woodroofe, M.\ 2009, \aj, 137, 3109 
\bibitem[Walker et al.(2009c)]{Waletal2009c} Walker, M.~G., Mateo, M., Olszewski, E.~W., et al.\ 2009, \apj, 704, 1274 
\bibitem[Walker \& Pe{\~n}arrubia(2011)]{WP2011} Walker, M.~G., \& Pe{\~n}arrubia, J.\ 2011, \apj, 742, 20 
\bibitem[Weisz et al.(2014)]{Weietal2014} Weisz, D.~R., Dolphin, A.~E., Skillman, E.~D., et al.\ 2014, \apj, 789, 147 
\bibitem[Weisz et al.(2015)]{Weietal2015} Weisz, D.~R., Dolphin, A.~E., Skillman, E.~D., et al.\ 2015, \apj, 804, 136 
\bibitem[Wolf et al.(2010)]{Woletal2010} Wolf, J., Martinez, G.~D., Bullock, J.~S., et al.\ 2010, \mnras, 406, 1220 
\end{thebibliography}
\end{document}